\documentclass[11pt, a4paper]{article}

\usepackage{fontawesome5}
\usepackage[T1]{fontenc} 

\usepackage{fontspec}

\usepackage[margin=2.0cm]{geometry} 
\usepackage{titling} 
\usepackage{graphicx}               
\usepackage{fancyhdr}               
\usepackage{titlesec}               
\usepackage{amsmath, amssymb, amsthm} 
\usepackage{enumitem}               
\usepackage{hyperref}               
\usepackage{url}                    
\usepackage[numbers,sort&compress]{natbib} 
\setcitestyle{square,comma} 
\usepackage{caption}                
\usepackage{subcaption}             
\usepackage{xcolor}                 
\usepackage{lipsum}                 
\usepackage{threeparttable}                 

\usepackage{tikz}   
\usetikzlibrary{positioning, fit, arrows.meta, shapes.geometric}

\usepackage{times}
\usepackage{booktabs}
\usepackage[utf8]{inputenc}
\usepackage[T1]{fontenc}
\usepackage{setspace}
\usepackage{microtype}
\usepackage[most]{tcolorbox}
\tcbuselibrary{breakable}

\usepackage{multirow}  
\usepackage{makecell}  
\usepackage{multicol}  

\hypersetup{
  colorlinks=true,
  linkcolor=blue,
  citecolor=blue,
  urlcolor=blue,
  linkbordercolor=white,
  citebordercolor=white,
  urlbordercolor=white
}

\fancypagestyle{firstpage}{
  \fancyhf{} 
  
  \chead{\raisebox{20pt}{\textbf{OxygenREC Technical Report}}} 

  \lhead{
    \begin{minipage}[b]{\textwidth} 
      \includegraphics[width=5cm]{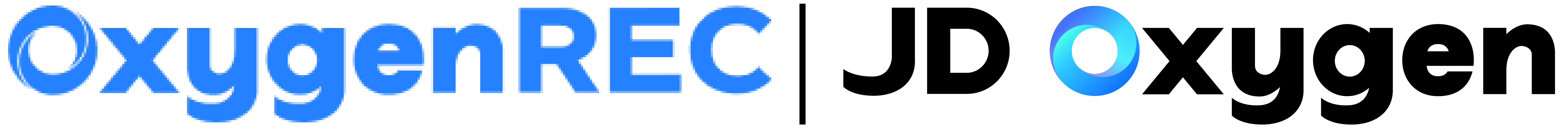}\\[-8pt] 
      \textcolor{black!70}{\rule{\textwidth}{1.0pt}}
    \end{minipage}
  }
  
  \rhead{\raisebox{20pt}{2025-12-31}}
  
  \cfoot{}
  \rfoot{\small \thepage}
  
  \setlength{\headheight}{60pt}
  \setlength{\topmargin}{-60pt} 

}

\fancypagestyle{fancy}{
    \fancyhf{} 

    \chead{\raisebox{32pt}{\textbf{OxygenREC Technical Report}}} 
    \lhead{} 
    
    \rhead{
        \smash{\raisebox{25pt}{ 
            \llap{\textcolor{black!70}{\rule{\textwidth}{1.0pt}}}
        }} 
    }
    \cfoot{} 
    \rfoot{\small \thepage} 

}
\titleformat{\section}[hang]
  {\fontsize{14pt}{16pt}\bfseries} 
  {\thesection.} 
  {0.5em} 
  {} 
  [] 

\titleformat{\subsection}[hang]
  {\fontsize{12pt}{14pt}\bfseries}
  {\thesubsection.}
  {0.5em}
  {}

\titleformat{\subsubsection}[hang]
  {\fontsize{11pt}{13pt}\itshape}
  {\thesubsubsection.}
  {0.5em}
  {}

\titlespacing*{\section}{0pt}{1.5em}{0.8em}
\titlespacing*{\subsection}{0pt}{1.2em}{0.6em}
\titlespacing*{\subsubsection}{0pt}{1em}{0.4em}

\usepackage{mathptmx}

\pretitle{%
  \vspace{-1.5cm}
  \fontsize{16pt}{18pt}\bfseries\raggedright\noindent
}
\posttitle{\par\vspace{-0.1em}}

\preauthor{\raggedright\fontsize{10pt}{12pt}\selectfont}
\postauthor{\par}

\title{OxygenREC: An Instruction-Following Generative Framework for E-commerce Recommendation}

\author{
    \fontsize{10pt}{12pt}\selectfont Xuegang Hao\textsuperscript{\faEnvelope}, Ming Zhang\textsuperscript{\faEnvelope}, Alex Li, Xiangyu Qian, Zhi Ma, Yanlong Zang, Shijie Yang, Zhongxuan Han, Xiaolong Ma, Jinguang Liu, Zhen Li, Zhida Jiang, Shusheng Wang, Ning Tang, Yanchen Qiao, Chenxiang Yang, Chen Sun, Jincheng Yuan, Chunhua Peng, Heng Hu, Peijun Yang, Baopeng Yuan, Caiyun Qiu, Zhaolong Xing, Haofei Yuan, Haipeng Zhang, Yuzhang Guo, Weijie Ding, Jiahua Gao, Hao Huang, Zhen Chen, Tongxuan Liu, Pinghua Gong\textsuperscript{\faEnvelope} \\
    \fontsize{8pt}{10pt}\selectfont JD.com, Beijing, China \space Contact: \{haoxuegang.1, zhangming229, gongpinghua1\}@jd.com
}

\date{} 

\begin{document}

\maketitle

\thispagestyle{firstpage}

\vspace*{-5.0em}

\begin{figure}[htbp]
    \centering
    \includegraphics[width=0.8\textwidth]{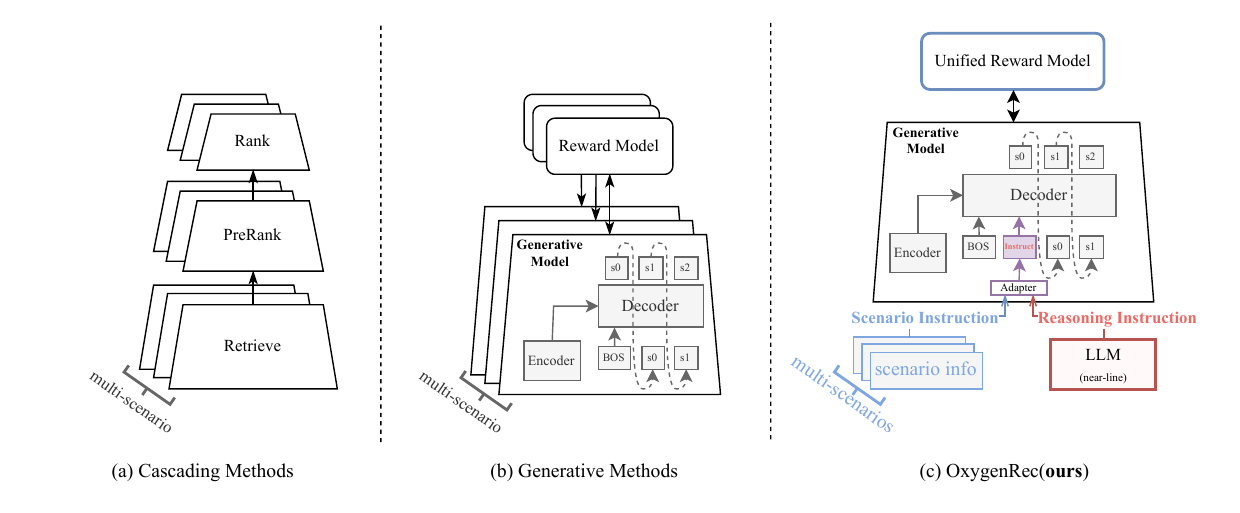} 
    \caption{Recommendation Architectures: (a) Cascading Methods: Fragmented optimization across multi stages. (b) Generative Methods: Unified model but limited by poor world-knowledge based reasoning and multi-scenario scalability. (c) OxygenREC (Ours): Achieves unification of deductive reasoning and multi-scenario adaption via Fast-Slow Thinking.}
    \label{fig:my_label}
\end{figure}

\noindent Traditional recommendation systems suffer from inconsistency in multi-stage optimization objectives. Generative Recommendation (GR) mitigates them through an end-to-end framework; however, existing methods still rely on matching mechanisms based on \textit{inductive patterns}. Although responsive, they lack the ability to uncover complex user intents that require deductive reasoning based on world knowledge. 
Meanwhile, LLMs show strong deep reasoning capabilities, but their latency and computational costs remain challenging for industrial applications. More critically, there are performance bottlenecks in multi-scenario scalability: as shown in Figure~\ref{fig:my_label}, existing solutions require independent training and deployment for each scenario, leading to low resource utilization and high maintenance costs---a challenge unaddressed in GR literature. To address these, we present OxygenREC, an industrial recommendation system that leverages Fast-Slow Thinking to deliver deep reasoning with strict latency and multi-scenario requirements of real-world environments.

\noindent First, we adopt a Fast-Slow Thinking architecture. \textit{Slow thinking} uses a near-line LLM pipeline to synthesize \textbf{Contextual Reasoning Instructions}, while \textit{fast thinking} employs a high-efficiency encoder--decoder backbone for real-time generation. Second, to ensure reasoning instructions effectively enhance recommendation generation, we introduce a \textit{semantic alignment mechanism} with \textbf{Instruction-Guided Retrieval (IGR)} to filter intent-relevant historical behaviors and use a \textbf{Query-to-Item (Q2I)} loss for instruction-item consistency. Finally, to resolve \textit{multi-scenario scalability}, we transform scenario information into controllable instructions, using unified reward mapping and \textbf{Soft Adaptive Group Clip Policy Optimization (SA-GCPO)} to align policies with diverse business objectives, realizing a \textit{``train-once-deploy-everywhere''} paradigm.

\noindent For large-scale deployment, we develop a unified PyTorch-based training framework that achieves $\mathbf{40\%}$ Model FLOPs Utilization (MFU), and the \textbf{xLLM}~\cite{xLLM} for high-performance inference serving, which has been open-sourced. OxygenREC achieves significant GMV and order volume increases in multiple core scenarios, demonstrating remarkable flexibility and scalability across diverse recommendation scenarios.

\vspace{-2.8ex}

\noindent\rule{4cm}{0.4pt} 
\vspace{-1.8ex}

{\footnotesize \textbf{JD Oxygen} is JD’s next-generation e-commerce AI architecture.}
\vspace{-3.0ex}

\clearpage
\setcounter{tocdepth}{2}
\tableofcontents

\clearpage

\section{Introduction}

Traditional recommendation systems usually use a multi-stage cascaded pipeline~\cite{cheng2016wide,zhu2025rankmixer,covington2016deep}, which often suffers from multi-stage objective misalignment and error propagation across stages. Recently, Generative Recommendation  (GR)~\cite{lin2025survey,zhou2025onerec} has emerged as a new paradigm that treats recommendation as an end-to-end sequence generation task, enabling global optimization and better efficiency. However, deploying GR in complex industrial settings still faces two key challenges:

\paragraph{First, limited deductive reasoning capabilities.} Current GR methods mostly learn inductively from user behavior~\cite{chen2025onesearch}. This falls short in scenarios that demand real-world knowledge and deep deductive reasoning. This manifests in two critical gaps:

(1) \textbf{Data Gaps in Complex Contexts: }
Traditional recommendation models depend on observed signals---such as location, time, and user demographics---but often struggle when encountering new or rarely seen combinations of context. For example, consider the scenario ``young parents in Chengdu during the winter solstice,'' represented by the triplet (Chengdu, winter solstice, young parents). Conventional systems typically fall back on generic suggestions like ``winter coats'', missing the real need of this couple: suitable clothing for their baby. In contrast, a \textit{Deductive Context-aware Reasoning} approach can infer---without relying on behavioral log---that ``moisture-wicking baby sleepwear'' better matches the user’s requirements, drawing on spatiotemporal knowledge about baby caring in cold-humid winter.

(2) \textbf{Logic Gaps in Intent Interpretation: }
Inductive models also face challenges when interpreting ambiguous user behavior. Imagine a photography enthusiast repeatedly comparing the Huawei Mate 70 and iPhone 16 Pro. Such activity is often misread as strong interest in both devices, leading to repetitive or irrelevant recommendations. However, the user might simply be unsatisfied because neither option fully meets his/her interest. In such case, a deductive recommender can clarify this ambiguity by incorporating additional context---for instance, the user’s history of watching photography tutorials---to deduce that high-quality mobile imaging is the true priority. As a result, it might suggest the Huawei Pura series---a phone designed for professional-level photography---even if the user has never viewed or interacted with it before.

Together, these examples show how deductive reasoning enhances GR systems beyond simple pattern recognition. By filling data gaps through commonsense generalization and resolving logic gaps via \textit{Context-Aware Intent Reasoning} and \textit{Deep Intent Disambiguation}, the system moves past statistical associations to act more like a knowledgeable assistant---one that truly understands what users need. Although integrating LLMs shows promise, directly using LLM backbones~\cite{he2025plum,chen2024hllm,liu2025onerec,zhou2025onerecv2} poses a critical trade-off: how to inject deductive world knowledge without causing unacceptably high online inference latency.

\paragraph{Second, the dilemma between multi-scenario adaptation and resource efficiency.} Industrial platforms deliver recommendation services across diverse scenarios (e.g., feeds and shopping-carts). While training dedicated models for each scenario brings additional operational and computational costs; conversely, a simple unified model often suffers from negative transfer issues. Most multi-scenario recommendation research primarily focuses on  \textit{discriminative} ranking models, relying on scenario-specific tower structures or complex routing/gating mechanisms to mitigate negative transfer~\cite{ma2018modeling,tang2020progressive,sheng2021one,caruana1997multitask}. In contrast, there are few works that focus on how to make generative recommendation work consistently across different types of scenarios.

To address the above challenges, we propose \textbf{OxygenREC}, an instruction-following unified generative framework for large-scale e-commerce recommendation. The proposed framework incorporates world knowledge for deductive intent reasoning and supports scalable multi-scenario services within a single unified backbone network. Our contributions are four-fold:

\begin{itemize}
    \item  \textbf{Fast-Slow Thinking Architecture with Deductive Knowledge Injection}. We propose a Fast-Slow Thinking architecture that injects world knowledge and deductive reasoning without introducing online latency: a near-line LLM pipeline performs \textit{slow} intent reasoning and generates high-precision \textit{Contextual Reasoning Instructions}, while a high-throughput encoder--decoder backbone network conducts \textit{fast} real-time generation based on these instructions.
    
    \item \textbf{Semantic Alignment for Effective Instruction Control}. To make instructions effective, we use a Query-to-Item (Q2I) loss to map them into the item embedding space. Since target items and user history share the same format, this mapping enables the instruction to search for relevant past items. We then use Instruction-Guided Retrieval (IGR) to filter out irrelevant history. By removing this noise, the model can focus solely on what the instruction asks for, ensuring the final output is tightly controlled by the user's intent.
    
    \item \textbf{Scalable Multi-Scenario Alignment via Instruction and Reinforcement Learning (RL)}. We convert scenario-specific contexts into scenario instructions within our generative model. To handle multiple business objectives, we use a unified Reward Mapping Service alongside Soft Adaptive Group Clip Policy Optimization (SA-GCPO) to sync a single policy with different business objectives. This approach allows us to ``train-once-deploy-everywhere'' while maintaining high performance across various scenarios.

    \item \textbf{Large-Scale Production Deployment}. We deployed OxygenREC in JD.com's core recommendation scenarios. Online A/B tests confirmed its impact, with significant gains in both order volume and GMV. The system is efficient and scales easily to handle massive traffic. It has proven to be a reliable engine for growth, even in the most demanding industrial environments.
\end{itemize}

The remainder of this paper is organized as follows. Section 2 presents the OxygenREC framework. Section 3 details the system implementation and optimization. Section 4 evaluates OxygenREC through comprehensive experiments. Section 5 concludes and outlines future work. For a detailed review of related work, see Appendix~\ref{app:related_work}.

\clearpage

\section{The OxygenREC Framework}

\subsection{Overview}

\begin{figure}[htbp]
    \centering
    \includegraphics[width=1.0\textwidth]{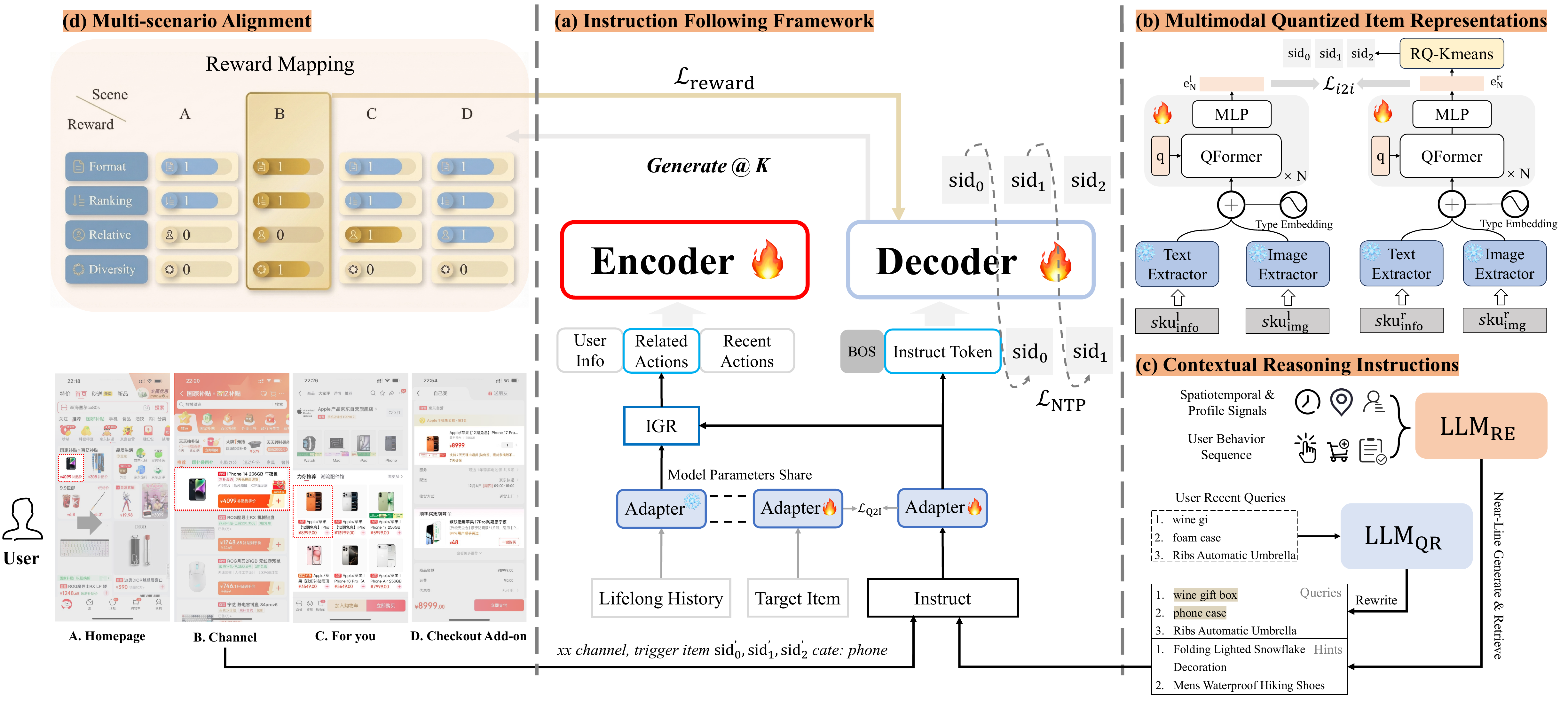} 
    \caption{\textbf{The Overall Architecture of OxygenREC.} 
    (a) \textbf{Instruction Following Framework:} A transformer-based encoder--decoder backbone that generates semantic item sequences conditioned on specific instructions.
    (b) \textbf{Multimodal Quantized Representations:} Items are tokenized as multimodal semantic IDs via residual quantization of contrastively trained embeddings, enabling compact and expressive item representations. 
    (c) \textbf{Contextual Reasoning Instructions:} A near-line LLM pipeline that analyzes user behavior and context to synthesize such instructions, bridging the gap between inductive patterns and deductive reasoning.
    (d) \textbf{Multi-Scenario Alignment:} We achieve a ``train-once-deploy-everywhere'' workflow by coupling scenario instructions with RL-based alignment.}

    \label{fig:overall_arch}
\end{figure}
OxygenREC unifies diverse recommendation tasks into a single \textbf{Instruction-Following Generative} paradigm. As illustrated in Figure~\ref{fig:overall_arch}, OxygenREC is designed to bridge the gap between high-level reasoning and real-time industrial recommendation through two key innovations, and addresses both challenges simultaneously:

\begin{itemize}
    \item \textbf{Fast-Slow Thinking:} A near-line LLM pipeline (\textit{slow thinking}) distills complex user intents into compact Contextual Reasoning Instructions, injecting world knowledge and deductive signals without incurring online LLM latency. A high-throughput encoder--decoder backbone (\textit{fast thinking}) then produces item sequences under strict latency constraints.
   \item \textbf{Low-Cost Multi-Scenario Adaptation:} Instead of separate models, we use one framework for all scenarios. Based on instruction control mechanism, we use scenario instructions to steer model behavior. For consistency, a reward mapping service and unified reward model provide a single standard to balance diverse objectives across the platform. During inference, prefix-constrained beam search enforces local business rules and candidate pools. This renders the system highly adaptable yet lean enough for large-scale deployment.
\end{itemize}

\noindent The end-to-end workflow consists of four stages:
\begin{enumerate}

    \item \textbf{Model Input Representation:} The input layer integrates three components into a unified space. User inputs consist of profile features and LLM-driven reasoning intents. Item inputs are represented by multimodal Semantic IDs (SIDs), which map items into a semantic latent space. Finally, contextual inputs provide real-time scenario information (see Appendix~\ref{app:scenario_instructions} for details), serving as the foundation for instruction control.
    \item \textbf{Instruction-Following Pre-training:} The model learns to follow instructions via a multi-task objective that combines Next Token Prediction (NTP) with semantic alignment.
    \item \textbf{Post-training with Multi-Scenario Alignment:} This stage leverages a reward mapping service to translate diverse, scenario-specific objectives (e.g., diversity, format) into unified reward signals, facilitating a novel policy optimization strategy for multi-objective refinement.
    \item \textbf{Multi-Scenario Serving:} This framework enables a single model to serve diverse scenarios simultaneously. It employs prefix-constrained decoding technology to strictly follow scenario-specific business rules and candidate pools during inference, thereby maximizing resource utilization efficiency.
\end{enumerate}

The subsequent sections detail these components: Section 2.2 introduces the SID and instruction construction; Section 2.3 describes the pre-training and controllable generation mechanisms; and Section 2.4 discusses post-training and multi-scenario alignment.

\subsection{LLM-driven User and Item Inputs}

\subsubsection{Model Input Representation}

As illustrated in Figure~\ref{fig:overall_arch} (a), OxygenREC adopts an encoder--decoder architecture. The encoder maps the user-side inputs into a latent space, while the decoder generates recommendation sequences conditioned on instructions. Below, we detail the input compositions for both components.

\begin{itemize}
    \item \textbf{Encoder Input ($X_{\text{enc}}$):} The encoder integrates three key input sources: (1) \textbf{User Profile}, capturing static attributes like demographics; (2) \textbf{User Behavior}, which is further split into short-term and long-term sequences. Short-term behaviors model the real-time evolution of user interests. For long-term history, we introduce IGR to filter a user's long-term history. By using the instruction as a query, the model retrieves only the past actions that are relevant to the current intent. This enhances instruction controllability, ensuring the model does not get distracted by old, irrelevant data. At the same time, it provides an efficient way to incorporate long-term interests, giving the model a deeper understanding of the user without the usual computational lag.

    \item \textbf{Decoder Input:} The decoder is conditioned on the encoded user representation $X_{\text{enc}}$ and a composite instruction prompt $P$. This prompt integrates two distinct signals: 
    (1) Scenario Instructions ($I_s$) for domain control, and (2) Contextual Reasoning Instructions ($I_r$) for deductive intent guidance.
    Together, they steer the auto-regressive generation of the target item sequence.
    
\end{itemize}

\subsubsection{Multimodal Quantized Item Representations}

As shown in Figure~\ref{fig:overall_arch} (b), we construct a unified vocabulary by training a multimodal item encoder and discretizing its output.
The encoder is optimized via a contrastive Item-to-Item (I2I) objective~\cite{radford2021learning,radford2018improving} using large-scale item pairs derived from cross-scenario co-occurrence behaviors. 
Each item is represented by textual metadata and product images, processed by separate encoders. 
To fuse these features, we employ a lightweight module where modality-specific projections map the inputs into a shared space, subsequently processed by Q-Former~\cite{li2023blip} and MLP layers to model cross-modal interactions.
The resulting 256-dimensional embeddings are then discretized using the RQ-KMeans scheme~\cite{gu2024residual}. This residual quantization process assigns each item a tuple of discrete codes in a coarse-to-fine manner. We employ a hierarchical structure with a depth of 3 and a vocabulary size of 8,192 per level, establishing a compact and expressive semantic ID space tailored for auto-regressive generation.

\subsubsection{Contextual Reasoning Instructions}
\label{sec:contextual_reasoning_instructions}

As illustrated in Figure~\ref{fig:overall_arch} (c), this section presents an architecture that integrates world knowledge and deductive reasoning while avoiding online LLM latency: a near-line LLM pipeline performs slow intent reasoning and synthesizes high-precision contextual reasoning instructions.  Our core goal is to leverage LLMs  to construct a controllable and interpretable intermediate instruction layer for generative recommendation systems. This instruction layer serves as an explicit semantic bridge between raw user behaviors and downstream retrieval or recommendation models, thereby improving intent alignment and system stability.

As shown in Figure~\ref{fig:LLMSFT}, we transform complex recommendation signals into a reasoning process based on multi-dimensional contexts. We build an instruction generation architecture with three parallel branches to map spatiotemporal context, search semantics, and historical behavior into text-based Instructions and corresponding Reasons. This explicit text modeling enhances the interpretability of recommendation decisions and allows us to monitor intermediate outputs independently.

\begin{figure}
    \centering
    \includegraphics[width=1\linewidth]{images/HintGenPipeline_v1.pdf}
    \caption{Overview of our Contextual Reasoning Instructions pipeline. Two parallel branches generate contextual instructions and reasons from spatiotemporal + profile signals and recent user behavior sequences. In parallel, $LLM_{QR}$ rewrites noisy or truncated recent queries, and all outputs are combined to form the final Instructions and Reasons for downstream generation}
    \label{fig:LLMSFT}
\end{figure}

\paragraph{Spatiotemporal and profile Reasoning (STP).}
This type of instruction focuses on inferring the user’s explicit needs under specific environment constraints. It covers three aspects: (i) Event-driven reasoning: it identifies important holidays or seasonal features in the current time and place, and infers related shopping intent. (ii) Profile-driven reasoning: it uses the user’s static traits (such as gender, age, and spending power) to form personalized product preference conclusions. (iii) STP Fusion reasoning: it combines local culture with real-time weather/time to infer localized needs, such as ``food intent related to the winter solstice in northern China''. See Section~\ref{app:prompt_design} in the supplementary material for a detailed case illustration.

In practice, we use the JoyAI LLM to generate these instructions and store them in a hierarchical way based on the ``time--location--persona'' triple. During online requests, we use the current time, location, and user profile as an index to retrieve and return results quickly.

\paragraph{User Query Rewrite.}
Real queries may have incomplete intent or semantic shift because of language noise or input truncation (such as ``foam case'' or ``wine gi''). We propose $LLM_{QR}$ to do semantic completion and normalization. Its training set includes two types of data: identity-preserving and rewrite-required. The core data comes from in-session self-correction traces, title alignment data guided by click feedback, and six types of synthetic error samples generated with DeepSeek-R1~\cite{guo2025deepseek}. We then perform supervised fine-tuning on Qwen3-0.6B~\cite{qwen3}. In human evaluation, this module reaches a pass rate of $95.33\%$, which is much better than the Vanilla DeepSeek-R1 baseline ($88.67\%$). This ensures high-quality conversion from the raw user need to normalized, intent-aligned instructions.

\paragraph{User Intent Reasoning.}

Real user searches usually carry clear and strong immediate intent. Given user behavior sequence and preference information, the model generates both intent-matched instructions and their reasoning reasons at the same time. Adding a rationale not only explains the logic behind the recommendation, but also helps the model find deeper motivations from noisy behaviors, which greatly improves intent recognition accuracy. See Section~\ref{app:prompt_design} in the supplementary material for a detailed case illustration.

To handle the lack of explicit rationale labels in real logs, we design a data refining and auto-labeling pipeline with multiple LLMs working together. First, we use $LLM_{QR}$ to map the final user query into a normalized query, and then use Qwen3-32B to aggregate and deduplicate it to get a standard intent target. Next, for the full behavior sequence, we use Qwen3-32B again to do alignment filtering, and keep only the subsequence that is strongly related to and supports the target intent, thereby reducing noise. Finally, based on the filtered sequence and the intent target, we use DeepSeek-R1 to automatically generate rationales as pseudo labels for training. After supervised fine-tuning on Qwen3-0.6B, the model generates intent-aligned instructions and reasons, and achieves a $72\%$ usability rate in human evaluation.

\subsection{Instruction-Following Unified Pre-training}

To support the goal of \textit{injecting deductive knowledge} and achieving \textit{multi-scenario adaptation}, the pre-training stage focuses on the following four components:

\begin{itemize}[leftmargin=*, noitemsep, topsep=0pt]
    \item \textbf{Dual-Instruction Formulation}: We define reasoning instructions ($I_r$) to inject deductive knowledge through a \textit{fast-slow thinking} mechanism, and scenario instructions ($I_s$) to adapt the model across diverse contexts.
    \item \textbf{Backbone with IGR}: This component strengthens the fast-slow logic by using intent from ``slow'' reasoning to filter the ``raw materials'' (user history) for the ``fast'' generator, improving both accuracy and efficiency.
    \item \textbf{Heterogeneous Data Mixture}: By combining search and diverse recommendation traffic, we provide varied instruction-label pairs. This facilitates \textit{instruction-following} generation and helps the backbone adapt to various scenarios across the app.
    \item \textbf{Multi-Objective Training}: We use joint learning objectives to align instructions with target items, supporting effective control of the generation process.
\end{itemize}

\subsubsection{Instruction Framework Design}
Our framework reformulates recommendation as an instruction-following generation task, $P(Y \mid X, I_s, I_r)$. By training the model to follow both scenario and reasoning instructions, the backbone learns to dynamically adjust its generation distribution according to the provided context.

\subsubsection{Dual-Instruction Formulation}
\label{sec:dual_instruction}

\paragraph{Composite Instruction Prompt.}
In practice, the instruction prompt consists of \textbf{Scenario Instructions} ($I_s$) and \textbf{Contextual Reasoning Instructions} ($I_r$). $I_s$ contains two fields: \textit{scenario information} and an optional \textit{trigger item}. $I_r$ is represented as a dense embedding projected from its textual source via the adapter. The availability of trigger items and the construction of $I_r$ vary across scenarios and between training and serving; we summarize these signals and the resulting conditional formulations in the pre-training section.

\paragraph{Scenario Instructions ($I_s$).}
$I_s$ specifies the scenario context for controllable generation. It includes (1) \textbf{scenario information} (available for all scenarios), and (2) an optional \textbf{trigger item} (available only for some scenarios). Scenario information may include a scenario ID as well as other contextual signals about the target scenario; it guides the generation style and the candidate/item distribution, since different scenarios may permit different item pools and users may exhibit different preferences across scenarios. When present, the trigger item provides local context (e.g., the channel entry item in Channel Feeds, or the main item for I2I related recommendations).
In practice, the trigger item is only available in scenarios such as channel feeds and I2I related recommendations, while scenarios like Homepage and Search typically do not have a trigger item.
Overall, $I_s$ helps one backbone serve multiple scenarios by adapting the generation style and candidate/item distribution, rather than training and deploying a separate model for each scenario.

\paragraph{Contextual Reasoning Instructions ($I_r$).}
$I_r$ is a \textit{dense embedding} obtained by projecting a \textit{textual instruction}---indicating users' latent intent through an adapter (Section~\ref{sec:architecture}). During online inference, the textual instruction is synthesized by our near-line LLM pipeline (Section~\ref{sec:contextual_reasoning_instructions} and Figure~\ref{fig:LLMSFT}). During training, to effectively enhance the model's instruction-following capability, we leverage the \textit{rewritten/normalized user query} in search data as a natural textual source for $I_r$; for recommendation scenarios where textual instructions are unavailable, we use a \textit{default (learnable) instruction embedding}. This design serves two purposes: (1) utilizing search queries as a natural and high-quality data source for instruction alignment, and (2) improving robustness in online serving, where near-line inference may fail or be delayed and $I_r$ may be missing---in such cases, the model can still produce reasonable recommendations conditioned on $I_s$ and user history.

\subsubsection{Generative Backbone with IGR}
\label{sec:architecture}
OxygenREC adopts an \textbf{encoder--decoder} architecture similar to OneRec~\cite{zhou2025onerec}, but distinguishes itself by specifically augmenting the decoder with the \textbf{Instruction Prompt} ($I_s, I_r$) as conditional inputs. This enables the model to auto-regressively generate item sequences that are explicitly guided by the user's latent intent and scenario context.

\noindent To enhance \textit{controllable generation}, we ensure that the user's historical context is consistent with the current instruction. Long-term user history typically contains many interactions that are irrelevant to the user's immediate intent. If the model processes the entire history, these irrelevant signals can interfere with its ability to follow the instruction. We therefore use IGR to filter the history and select interactions that are most relevant to the instruction prompt. This ensures that the input context is aligned with the control signals ($I_s, I_r$), leading to more precise output. As illustrated in Figure~\ref{fig:overall_arch} (a), this mechanism consists of three components: (1) \textbf{Adapter}: projects instructions and items into a shared embedding space; (2) \textbf{Q2I alignment}: uses the ground-truth target to supervise instruction--item similarity during training; and (3) \textbf{IGR}: performs top-$K$ retrieval at inference time to provide the decoder with an instruction-aligned context.

\paragraph{Adapter Mechanism for Feature Mapping.}
For similarity-based retrieval, the query and history items must be comparable in the same space. However, the query is driven by $(I_s, I_r)$, while history items are described by item IDs, side features, and texts. We use adapter layers~\cite{houlsby2019parameter} to project them into a shared embedding space. The textual instruction $I_r^{\text{text}}$ is generated by the near-line pipeline in Section~\ref{sec:contextual_reasoning_instructions} and is encoded by the same text encoder $g(\cdot)$ used for item texts. The embeddings are defined as:
\begin{equation}
\begin{aligned}
\mathbf{e}_q &= \text{Concat}\left[\phi_{\text{scn}}(I_s), g^{\text{train}}(I_r^{\text{text}})\right] \\
\mathbf{e}_t &= \text{Concat}\left[\phi_{\text{item}}(v_t), \phi_{\text{side}}(u_t), g^{\text{train}}(x_t)\right] \\
\mathbf{e}_h &= \text{Concat}\left[\phi_{\text{item}}(v_h), \phi_{\text{side}}(u_h), g^{\text{frozen}}(x_h)\right] \\
\mathbf{q} &= \psi_{q}(\mathbf{e}_q), \quad \mathbf{t} = \psi_{i}(\mathbf{e}_t), \quad \mathbf{h} = \psi_{i}(\mathbf{e}_h)
\end{aligned}
\end{equation}

where $\psi_{q}$ and $\psi_{i}$ are projection networks. $I_s$ consists of scenario information $s$ and trigger item $z$ (using $z=z_{\text{def}}$ when absent). $\phi_{\text{scn}}(\cdot)$ projects scenario information into the embedding space. $\phi_{\text{item}}(\cdot)$ and $\phi_{\text{side}}(\cdot)$ embed item IDs ($v_t, v_h$) and side-information features ($u_t, u_h$). $x_t$ and $x_h$ are the textual descriptions of the target item and history items. $g(\cdot)$ is the shared text encoder; we denote trainable/frozen usage as $g^{\text{train}}(\cdot)$ and $g^{\text{frozen}}(\cdot)$. Since long-term histories can be very long, we do not backpropagate gradients through the long-history branch (including $g^{\text{frozen}}(x_h)$) during training to reduce compute and communication overhead.

\paragraph{Q2I Alignment.}
To make the query embedding $q$ and history item embeddings $\{h\}$ comparable for retrieval, we align $q$ with the \textit{target item embedding} $t$. The target item $t$ is available only during training and serves as a supervised signal to anchor the query into the item space. Since both $t$ and history items are projected by the same adapter $\psi_{i}$, this alignment ensures that $q$ can be used to retrieve relevant history interactions at serving time when the ground-truth target is absent. Specifically, for a batch of size $B$ with normalized query embeddings $Q = \{q_1,...,q_B\}$ and normalized target item embeddings $T = \{t_1,...,t_B\}$, we optimize the following auxiliary objective:
\begin{equation}
\mathcal{L}_{\text{Q2I}} = \underbrace{-\frac{1}{B}\sum_{i=1}^B q_i \cdot t_i}_{\text{Alignment}} + \lambda_r \underbrace{\left(-\log\left[\text{Var}(\mathbf{Q}) \cdot \text{Var}(\mathbf{T})\right]\right)}_{\text{Regularization}} + \lambda_d \underbrace{\frac{1}{B^2-B}\sum_{i\neq j} (q_i^\top q_j)^2}_{\text{Decorrelation}}
\end{equation}
where $\lambda_r$ and $\lambda_d$ control the strengths of the regularization and decorrelation terms. $\text{Var}(\mathbf{Q})$ and $\text{Var}(\mathbf{T})$ denote the average variance (across embedding dimensions) within the batch. The regularization and decorrelation terms help avoid dimensional collapse and reduce embedding redundancy~\cite{pmlr-v139-zbontar21a,2021Understanding}.

\paragraph{IGR.}
With the aligned embedding space, IGR uses the query embedding $q$ to retrieve the top-$K$ most relevant interactions from the long-term history (in the $\psi_{i}$ space). This does more than just reduce noise and shorten input sequences for efficiency; it directly strengthens instruction control. By filtering out irrelevant history, IGR ensures the model stays locked onto the user's current request rather than being distracted by past habits.

\subsubsection{Instruction-Following Pre-training: Data Mixture, Signals, and Objectives}
\label{sec:objectives}

\paragraph{Data Mixture.}

Our pre-training data is a mixture of \textit{search} and \textit{multiple recommendation scenarios} (e.g., Homepage, Channel Feeds, and I2I related recommendations). This design brings several benefits. First, it enables unified modeling of user trajectories across the app, since real user journeys often interleave browsing, channel entry, item-to-item exploration, and search. Second, it substantially expands the amount of usable supervision: in e-commerce, positive feedback signals (e.g., purchase and add-to-cart) are typically sparser than in content platforms, and incorporating both search and recommendation traffic improves coverage and data efficiency. Finally, training on scenarios with and without contextual reasoning instructions improves robustness when the instruction is missing in online serving.

\paragraph{Training Signals and Scenario-Specific Formulation.}
Across all scenarios, we always include \textit{scenario information} as part of $I_s$. Some scenarios additionally provide an optional \textit{trigger item} (e.g., channel entry item or the main item on product detail page/cart). For $I_r$, only search data provides an observed query as a direct textual source; for recommendation scenarios where the near-line instruction is unavailable during training, we use a \textit{default (learnable) instruction embedding}. Table~\ref{tab:scenario_instructions} summarizes the resulting training signals and scenario-specific conditional formulations.
\begin{table}[htbp]
\centering
\footnotesize
\setlength{\tabcolsep}{4pt}
\begin{tabular}{@{}p{3.1cm}cccc@{}}
\toprule
\textbf{Scenario} & \textbf{Scenario Info} & \textbf{Trigger Item} & \textbf{Contextual Reasoning} & \textbf{Formulation} \\
\midrule
Search & \checkmark & $z_{\text{def}}$ & \makecell{query-derived} & $P(Y \mid X, I_s(s, z_{\text{def}}), I_r(q))$ \\
Homepage & \checkmark & $z_{\text{def}}$ & \makecell{default emb.} & $P(Y \mid X, I_s(s, z_{\text{def}}), I_r^{\text{def}})$ \\
Channel Feeds & \checkmark & $z_{\text{entry}}$ & \makecell{default emb.} & $P(Y \mid X, I_s(s, z_{\text{entry}}), I_r^{\text{def}})$ \\
\makecell[l]{Related Rec.\\(Item-to-Item)} & \checkmark & $z_{\text{main}}$ & \makecell{default emb.} & $P(Y \mid X, I_s(s, z_{\text{main}}), I_r^{\text{def}})$ \\
\bottomrule
\end{tabular}
\caption{Training signals and scenario-specific conditional formulations. Scenario information ($s$) is always available. The trigger item is denoted by $z$ (using $z_{\text{def}}$ when absent). $I_r(q)$ denotes the contextual reasoning instruction embedding projected from the (rewritten/normalized) query, while $I_r^{\text{def}}$ denotes a default learnable embedding used when the textual instruction is unavailable in training.}
\label{tab:scenario_instructions}
\end{table}

\paragraph{Joint Learning Objectives.}
The training objective combines generation accuracy with query-to-item alignment:
\begin{equation}
\mathcal{L} = \mathcal{L}_{\text{NTP}} + \lambda \mathcal{L}_{\text{Q2I}}
\end{equation}

\paragraph{Weighted Next Token Prediction ($\mathcal{L}_{\text{NTP}}$).} To prioritize high-value user behaviors, we optimize the auto-regressive likelihood of the target item sequence with a weighted NTP loss. Higher weights are assigned to conversion-related tokens (e.g., Purchase > Cart > Click).

\subsection{Post-training with Multi-Scenario Alignment}

\subsubsection{Post-training Framework}
In the post-training phase, unlike previous methods that perform fine-tuning in a single scenario, we adopt a multi-scenario modeling approach. OxygenREC post-training primarily consists of two components: a reward mapping system based on different scenarios to provide feedback signals, and a post-training process based on RL designed for various tasks across different scenarios. We construct a real-online reward mapping system that includes an online unified ranking model service and other multi-task rewards. We propose SA-GCPO to adopt a soft-adaptive function to compute importance sampling weights while incorporating reward scores derived from real user behavior as the threshold to distinguish positive and negative advantage samples.

\begin{figure}[t]
    \centering
    \includegraphics[width=0.8\textwidth]{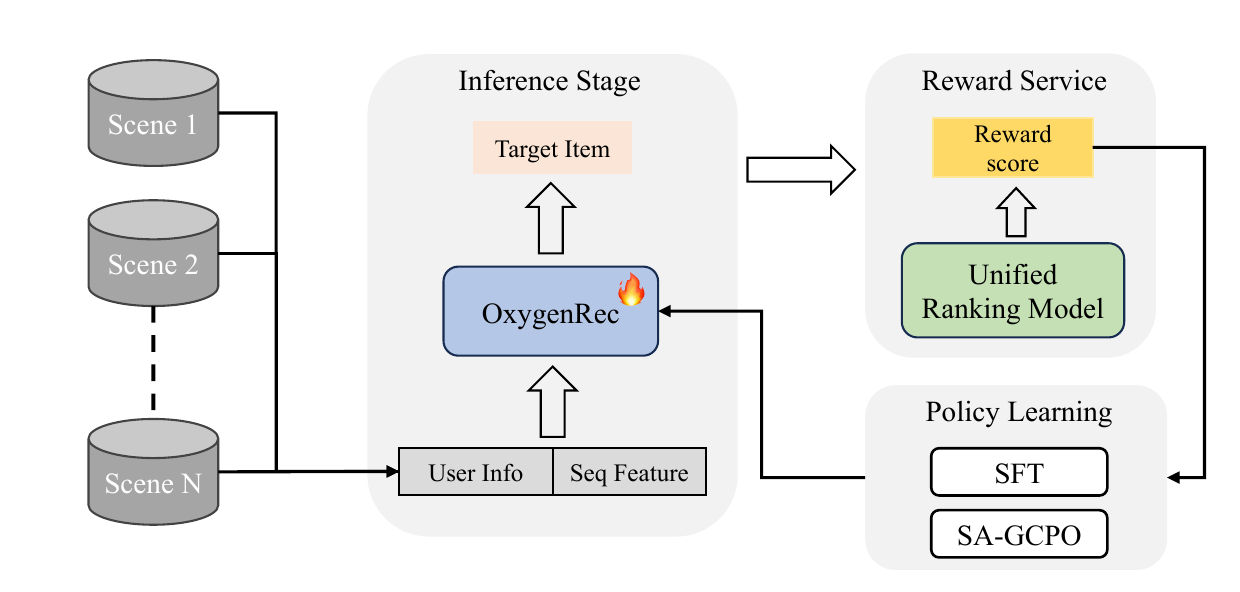} 
    \caption{The post-training process of OxygenREC}
    \label{fig:post_training_process}
\end{figure}

The post-training process of the OxygenREC primarily consists of three modules: the reward service, inference stage, and policy learning, as illustrated in Figure~\ref{fig:post_training_process}. For multi-scenario information, we construct features based on the user’s historical profile and sequential behavior data across multiple scenarios, which are then fed into the policy model for generation. During the inference stage, multiple candidate items are generated. We employ the unified ranking model as an online reward service, which returns a score based on the requested item and user information via an online ranking service and other rule-based methods. Finally, we conduct SFT and RL for policy learning. The policy model is updated and we repeat this process for the next iteration until the model converges.

\subsubsection{Multi-Scenario Adaptation}
Most current GR methods ~\cite{zhou2025onerec,zhou2025onerecv2,chen2025onesearch,liu2025onerec} conduct SFT and RL separately for different scenarios during post-training stage. These scenario-specific training processes are then paired with a ranking model that provides reward signals for RL. However, this approach necessitates the costly development and online deployment of individual GR model and ranking service for each scenario. To address this inefficiency, our reward mapping system employs a unified ranking model as a centralized reward model service. Then we conduct the post-training collectively on data from various scenarios, which significantly reduces the computational costs associated with both training and large-scale online deployment. To achieve robust performance across diverse scenarios, we construct a scenario-aware reward mapping as follows:

\paragraph{Scenario-Aware Reward Mapping.} As shown in Figure~\ref{fig:overall_arch}~(d), the system incorporates RL stage using rewards tailored to specific recommendation scenarios. The total reward score is a weighted combination of following distinct rewards:
\begin{enumerate}
    \item \textbf{Format Reward:} Penalize structural errors to ensure valid item semantic ID output.
    \item \textbf{Relative Reward:} Reward items relevant to the user's immediate context and query (based on Q2I semantic relationships).
    \item \textbf{Ranking Reward:} Reward sequences that maximize business objectives like GMV or Conversion Rate.
    \item \textbf{Diversity Reward:} Evaluate the diversity of a group of items generated by the GR model.
\end{enumerate}

\begin{figure}[t]
    \centering
    \includegraphics[width=1.0\textwidth]{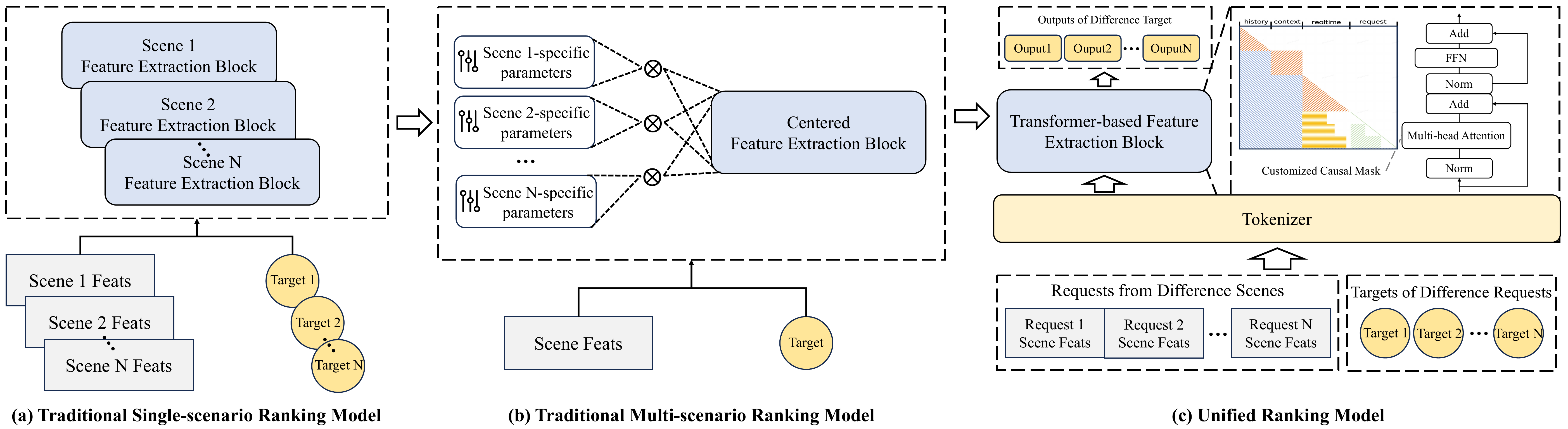} 
    \caption{Comparison of model architectures: traditional ranking models vs. our unified ranking model}
    \label{fig:unifiedranking}
\end{figure}

\paragraph{Unified Ranking Model.}
We develop a novel unified multi-scenario ranking model to serve as the core of the Reward Mapping Service, providing a consistent ranking reward across diverse scenarios.  Many traditional multi-scenario recommendations, such as STAR~\cite{sheng2021one} and PEPNet~\cite{chang2023pepnet}, introduce complex specialized structures to mitigate the negative transfer problem and require maintaining separate model instances or intricate routing logic, leading to high operational and maintenance costs. To address these problems in GR, we propose a unified list-wise approach for both offline training and online learning of our ranking models. By constructing unified representation tokens from heterogeneous user features and processing them through a shared Transformer-based feature extraction block, we achieve comprehensive cross-scenario feature interaction and consistent scaling effects, as shown in Figure~\ref{fig:unifiedranking}. The model further incorporates an adaptive modeling mechanism designed to address input feature heterogeneity, thereby comprehensively enhancing its representational capacity.

We construct multi-scenario training samples by using a label packing strategy, transforming traditional point-wise samples into list-wise samples. These samples are arranged chronologically according to users' request sequences, with a customized causal masking mechanism applied during training to explicitly model user behavior trajectories. Through explicit behavior trajectory modeling, the system can maximize conversion gains throughout the entire user path while improving the efficiency of joint recommendations across multiple positions. Compared to existing methods, the proposed unified ranking model demonstrates more consistent scaling effects across different scenarios and tasks.

\subsubsection{Reinforcement Learning }

Preference alignment during post-training is critical for enhancing model performance. Numerous studies~\cite{guo2025deepseek} adopt GRPO for policy learning. OneRec-v1~\cite{zhou2025onerec} proposes ECPO, which clips policies with large ratios to ensure training stability. OneRec-v2~\cite{zhou2025onerecv2} truncates gradients within a bound to address instability arising from negative samples. However, in our scenario, the issue of training instability becomes more pronounced as OxygenREC requires unified training across multiple environments. In addition, because our reward mapping involves scores from various tasks such as format, relative, and ranking reward, the hard clipping strategy adopted by most existing methods often leads to sample inefficiency and discontinuous gradients. Recently, several works~\cite{gao2025soft} in language modeling have employed smooth gate functions rather than hard clipping to maintain training stability. By utilizing soft gate functions, gradients can decay smoothly, preserving more learning signals while ensuring stability. 

\paragraph{Soft Adaptive Group Clip Policy Optimization.} 
In this section, we propose SA-GCPO, adopting a soft adaptive method to compute importance sampling weight, while also incorporating reward scores derived from real user behavior as the threshold to distinguish positive and negative advantage samples. We further adopt an asymmetric temperature control mechanism, applying different temperature coefficients to positive and negative advantage samples, respectively. This design accelerates the gradient decay of negative samples, thereby alleviating issues of gradient diffusion and training instability. Unlike~\cite{gao2025soft}, which employs the sign of advantage values to distinguish between positive and negative advantage samples, we propose a method based on real user feedback to define them. In most RLVR methods~\cite{guo2025deepseek,team2025kimi,lambert2024tulu} for LLM reasoning and mathematical tasks, rewards often have clear positive or negative definitions. However, due to the multi-scenario and multi-task nature of our reward design in OxygenREC, simply relying on the sign of the advantage value becomes insufficient. To address this, we introduce reward scores derived from real user feedback as a threshold to define positive and negative advantages, and subsequently apply different temperature coefficients. This approach enables the explicit definition of positive feedback samples during such multi-scenario, multi-task post-training. We summarize our proposed SA-GCPO as follows:

Given each sample $x \sim \mathit{D}$, for a group of items $\{y_i\}_{i=1}^G$ generated from behavior policy $\pi_{\theta_{\text{old}}} $ for sample $x$, our proposed SA-GCPO employs the following  optimization objective:

\begin{equation}
\mathcal{J}_{\text{SA-GCPO}}(\theta) = \mathbb{E}_{x \sim \mathit{D},\{y_i\}_{i=1}^G \sim \pi_{\theta_{\text{old}}}(\cdot|x)} 
\left[ \frac{1}{G} \sum_{i=1}^{G} \frac{1}{|y_i|} \sum_{t=1}^{|y_i|} f_{i,t}(r_{i,t}(\theta)) \Gamma^{\text{adv}}(\widehat{A}_{i,t}, R^{*}_g) \right],
 \label{equ_obj}
\end{equation}
where $G$ is the number of generated items for each sample $x$ and $r_{i,t}(\theta)$ represents the token-level importance ratio. The definition is as follows:
    \begin{equation}
    r_{i,t}(\theta) = \frac{\pi_{\theta}(y_{i,t} \mid x, y_{i,<t})}{\pi_{\theta_{\text{old}}}(y_{i,t} \mid x, y_{i,<t})}
    \end{equation}
In Equation~\ref{equ_obj}, $f_{i,t}(\rho)$ denotes the soft adaptive function: 
    \begin{equation}
    f_{i,t}(\rho) = \sigma\left( \tau_{i,t} (\rho - 1) \right) \cdot \frac{4}{\tau_{i,t}}, \ \ \
    \tau_{i,t} = 
    \begin{cases}
    \tau_{\text{pos}}, & \Gamma^{\text{adv}}(\widehat{A}_{i,t}, R^{*}_g) > 0, \\
    \tau_{\text{neg}}, & \text{otherwise},
    \end{cases}
    \label{equ4}
    \end{equation}
where $\tau_{\text{pos}}$ and $\tau_{\text{neg}}$ are the temperatures set for positive and negative samples. $\sigma(z) = {1}/({1 + e^{-z}})$ is the sigmoid function. Its gradient weight $w_{i,t}(\theta)$ is given by:
    \begin{equation}
    w_{i,t}(\theta) = 4 \, p_{i,t}(\theta) \left( 1 - p_{i,t}(\theta) \right), \quad p_{i,t}(\theta) = \sigma\left( \tau_{i,t} \left( r_{i,t}(\theta) - 1 \right) \right),
    \end{equation}
which peaks at $ r_{i,t}(\theta) = 1 $ and decays smoothly as $r_{i,t}(\theta)$ deviates, implementing a soft trust region of $r_{i,t}(\theta)$. 
\newline $\Gamma^{\text{adv}}$ in Equation~\ref{equ_obj} represents the threshold function to distinct the positive and negative advantage samples of $\widehat{A}_{i,t}$, which is denoted as follows:
    \begin{equation}
     \Gamma^{\text{adv}}(\widehat{A}_{i,t}, R^{*}_g) = 
    \begin{cases}
    0, & \widehat{A}_{i,t} > 0 \text{ and } R_i<R^{*}_g, \\
    \widehat{A}_{i,t}, & \text{otherwise},
    \end{cases}
    \end{equation}
where $R_i$ is the reward score for the item $y_i$ and $R^*_g$ denotes the reward of the target item in this group. $\widehat{A}_i$ is the normalized advantage for item $y_i$, the calculation of $\widehat{A}_i$  given by:
    \begin{equation}
     \widehat{A}_{i,t} =\widehat{A}_i = \frac{R_i - \text{mean}(\{R_i\}_{i=1}^G)}{\text{std}(\{R_i\}_{i=1}^G)}
\end{equation}

The main advantages of the design in our methods are as follows:

\begin{itemize}
    \item \textbf{Adaptive smooth gating:} Replace hard clipping function with a continuous sigmoid-based gate function, reducing optimization noise and enhancing training stability.
    \item \textbf{Real user feedback as threshold for pos/neg advantages:} The reward scores derived from user real feedback are employed as the threshold for distinguishing positive and negative advantage samples, thereby mitigating the issue of reward hacking during the RL training process.
    
    \item \textbf{Asymmetric temperature control:} Uses a different temperature setting for $\tau_{\text{neg}}$ and $\tau_{\text{pos}}$ to more rapidly attenuate negative-token gradients, improving stability.    
    \item \textbf{Sequence-level coherence:} Since the SID represents a single item, SA-GCPO reduces to a smooth sequence-level gate similar to GSPO~\cite{zheng2025group} but without abrupt clipping.

\end{itemize}

\section{System Implementation and Optimization}

\subsection{Unified Training Framework}
The training of OxygenREC presents a unique engineering challenge, which requires simultaneously handling \textit{terabyte-scale sparse embeddings} (typical of recommendation systems) and \textit{billion-scale dense parameters} (typical of LLMs). Traditional frameworks create a divide: TensorFlow excels at sparse optimization but lacks mature LLM support, while PyTorch dominates LLM training but it is not applicable with industrial-scale embeddings.

To bridge this gap, we propose a unified training framework built on PyTorch~\cite{li2020pytorch} that integrates industrial-grade sparse and dense engines. Inspired by efficient distributed training techniques~\cite{shoeybi2019megatron,huang2019gpipe}, the training is conducted on a production cluster with 128 NVIDIA H800 GPUs, utilizing 400GB/s NVLink and 8×200Gb/s RoCEv2 RDMA for high-performance interconnects. We achieve an overall Model FLOPs Utilization (MFU) of 40\% through the following key optimizations.

\subsubsection{Distributed Sparse Optimization}
We designed a large-scale distributed sparse engine to handle massive embedding tables efficiently within PyTorch. It implements a non-overlapping partition strategy for embeddings across workers, coupled with hierarchical HBM-MEM caching. A multi-stage pipeline is established to hide the inherent latency of embedding access. To avoid embedding staleness, we implement a dual-buffer mechanism to provide strong consistency guarantees. Our optimizations reduce the time proportion of sparse operations from 15\% to 5\% and achieve 1.1-2.4$\times$ speedup over state-of-the-art open-source embedding solutions.

\subsubsection{Operator-Level Acceleration}
For the LLM backbone, we integrate mature optimizations including BF16 mixed-precision training~\cite{kaplan2020scaling} and ZeRO~\cite{rajbhandari2020zero} for memory efficiency. We also employ advanced attention mechanisms~\cite{dao2022flashattention,shazeer2020glu} and efficient architectures~\cite{fedus2022switch,sha1345}. However, standard attention implementations (e.g., FlashAttention) are not suitable for the hybrid masking patterns required in generative recommendation. To address this, we develop a dedicated attention acceleration library. By leveraging CUTLASS~\cite{Thakkar_CUTLASS_2023} and TileLang~\cite{TileLang} for custom kernel compilation, our training framework supports flexible mask configurations and achieves speedups of \textbf{1.7$\times$ and 3.0$\times$} compared to FlexAttention~\cite{dongflexattention} and torch.compile~\cite{ansel2024pytorch}, respectively.

\subsubsection{Scenario-Aware Reinforcement Learning}
The post-training RL phase involves distinct workflow designs and model characteristics, particularly with respect to handling large-scale sparse parameters and multi-scenario customization. We build a customized RL workflow on top of Ray~\cite{moritz2018ray}. In collocated deployment modes, we enable shared-memory access for sparse tables to eliminate redundant copying, ensuring efficient synchronization during the high-throughput sample generation phase.

\subsection{Inference Optimization based on xLLM}
GR integrates LLMs to enhance sequence understanding, but its inference workload differs markedly from standard LLM serving. GR typically processes long user-history prompts while generating relatively short outputs under a large beam width (e.g., beam size = 256-512). This setting makes decoding the dominant bottleneck: beyond the increased number of active hypotheses, large-beam decoding incurs substantial extra overhead in candidate sorting and stochastic sampling, and it amplifies KV-cache capacity/traffic pressure and memory access inefficiency compared to standard greedy sampling.

To meet strict Service Level Objectives (SLOs), we develop \textbf{xGR}~\cite{xGR}, a dedicated serving system built upon the high-performance \textbf{xLLM} framework~\cite{xLLM}. xGR adopts a tightly coupled \textit{three-tier architecture} to maximize hardware efficiency:
\begin{itemize}
    \item \textbf{xSchedule (System Level):} A sophisticated scheduler that manages task parallelism. It enables fine-grained pipeline overlapping across batching, request handling, and kernel execution, ensuring high GPU utilization.
    \item \textbf{xAttention (Operator Level):} Built upon xLLM's PagedAttention, this module is further customized for GR's attention patterns (e.g., long prompts + short decoding, hybrid masks). It strengthens \textit{KV-cache management} and introduces \textit{staged compute allocation} to better match the large-beam decoding workload.
    \item \textbf{xBeam (Algorithm Level):} A specialized module designed to handle the massive sorting overhead of large-beam decoding and support advanced sampling strategies over billion-scale item spaces.
\end{itemize}

\paragraph{Deep Customization for OxygenREC.}
Leveraging this architecture, We implement critical optimizations for OxygenREC:

\textbf{Specialized Beam Search \& Sampling.} Unlike standard NLP tasks that prioritize deterministic outputs, recommendation requires balancing accuracy with diversity. xBeam implements a highly optimized \textit{Beam Sample} kernel that efficiently combines top-k selection with nucleus/multinomial sampling. It employs operator-level fusion to handle the stochasticity without performance degradation, significantly reducing decoding latency compared to vanilla implementations.

\textbf{Prefix-Constrained Decoding.} To enforce scenario-specific rules, we integrated a \textit{Trie Index} mechanism into the inference loop. This lightweight index dynamically generates logit masks at each step, guaranteeing 100\% generation legality within designated item pools with negligible runtime overhead.

\subsection{Inference Deployment of Reasoning Instructions}
\label{subsec:deployment}

We adopt a \textit{near-line updating mechanism} to achieve a balance between the timeliness of recommendation signals and overall system overhead. Our deployment architecture consists of two cascaded components: an LLM-based instruction generation service and an Adapter-based text encoder service.

The LLM instruction model operates in a near-line setting, where it synthesizes natural-language \textit{reasoning instructions} by fusing spatiotemporal context and user behavioral history. These instructions are not generated on-the-fly during user requests; instead, they are produced in batch. The subsequent Adapter text encoder then converts each textual instruction into a dense embedding vector. This embedding is indexed by the user ID and stored in a low-latency key-value store (e.g., Redis cluster), ready for real-time consumption by the downstream generative recommender.

To capture both long-term preferences and short-term intent shifts, we implement a hybrid update strategy:

\begin{enumerate}[leftmargin=*,nosep]
    \item \textbf{Daily full refresh}: An offline job regenerates spatiotemporal and behavioral instructions for all daily active users, incorporating the latest contextual signals (e.g., season, location) and accumulated interactions.
    
    \item \textbf{Near-line incremental update}: When a user performs high-value actions---such as effective searches, views, cart additions, or purchases---the system triggers a near-real-time instruction update. To prevent excessive write pressure and system instability caused by bursty interactions, we employ a \textit{time-window aggregation} strategy: within a short sliding window (e.g., 5 minutes), multiple behavioral events from the same user are merged into a unified intent summary, and only a single instruction regeneration and storage write is executed at the window’s end. This design preserves signal responsiveness while significantly reducing backend load.
\end{enumerate}

During online inference, the generative recommendation model simply retrieves the precomputed instruction embedding via the user ID---enabling \textit{zero online LLM calls}, \textit{low-latency serving}, and \textit{high semantic fidelity} in the final recommendations.

\section{Evaluation and Results}

\subsection{Evaluation Metrics}

\paragraph{Model Evaluation Metrics.}

\begin{itemize}
    \item \textbf{HitRate@K (HR@K):} This metric quantifies the precision of the generative process by measuring the proportion of test instances where the generated candidate strictly aligns with the ground-truth semantic ID sequence across all hierarchical code levels within the top-$K$ beam search hypotheses~\cite{freitag2017beam}.
    
    \item \textbf{Recall@K:} This metric assesses the model's capacity to cover the user's relevant interests. It is calculated as the ratio of the user's daily positive interactions that are successfully identified within the top-$K$ generated candidates relative to the full set of observed user behaviors for that day.

\end{itemize}

\paragraph{Semantic ID Evaluation Metrics.}

\begin{itemize}
    \item \textbf{SID Codebook Coverage:} This metric evaluates the utilization efficiency of the hierarchical semantic space. Unlike simple code usage, it measures the \textit{joint path occupancy} at each quantization level. Specifically, for a given depth $L$ (where $L \in \{1, 2, 3\}$), the coverage is defined as the ratio of unique code tuples $(c_1, \dots, c_L)$ actually assigned to the active SKU set relative to the total theoretical capacity of that conceptual space (i.e., $|V|^L$, where $|V|$ is the codebook width). This reflects how effectively the SKUs populate the exponentially growing combination space.

    \item \textbf{Semantic Cluster Purity:} This metric evaluates the alignment between the learned semantic clusters and human-defined taxonomies. It is calculated by measuring the mean dominance of the primary category within each semantic ID's corresponding SKU set. A higher purity score indicates that the semantic IDs effectively capture the high-level category semantics (e.g., SKUs grouped under a specific semantic ID consistently belong to the ``Electronics'' category rather than a mix of unrelated categories).

    \item \textbf{Semantic ID Collision:} This metric quantifies the discriminative power of the semantic identifiers in distinguishing individual items within the billion-scale SKU pool. It is measured by the distribution of item cardinality for each unique SID tuple. We report the SKU counts at key quantiles (e.g., P90, P99). A lower SKU count at these upper quantiles indicates finer granularity, meaning the semantic ID sequence provides a more precise representation closer to a unique item identifier.

    \item \textbf{Codebook Load Balance:} This metric measures the uniformity of item distribution across the quantization codebook. It quantifies the deviation of actual cluster sizes (number of SKUs assigned to a specific code) from the theoretical uniform distribution (Total SKUs / Codebook Size). By examining this ratio at various quantiles, we assess whether the RQ-KMeans process effectively utilizes the full capacity of the codebook or suffers from mode collapse (where specific codes are over-utilized while others remain under-utilized).
\end{itemize}

\subsection{Ablation Studies and Analysis}

In this section, we conduct a comprehensive ablation study to validate the effectiveness of our proposed components. We first analyze the evolution of SID representations and their impact on retrieval quality. Subsequently, we explore the scalability and performance of the generative backbone under various architectural configurations.

\paragraph{Lightweight Multimodal Design.}
We conduct a systematic analysis of the multimodal encoder design by progressively ablating and iteratively refining its architecture. This process starts from a faithful reproduction of the OneRec-style multimodal model, which adopts a large pretrained multimodal backbone (e.g., MiniCPM-V~\cite{minicpmv}) followed by multiple stacked Q-Former~\cite{li2023blip} layers for feature compression.

To reduce computational cost, we first replace the heavy multimodal backbone with a series of lighter vision-language models, including variants from the SAIL~\cite{chen2025sail} and InternVL~\cite{chen2024internvl} families. Although inference becomes significantly faster, the overall recommendation performance shows almost no improvement. We further perform a systematic comparison among pure-text encoders, pure-vision encoders, and multimodal backbone models~\cite{wang2023one,li2024llava,liu2024mmgrec}. Surprisingly, the pure-text encoder Qwen3~\cite{qwen3,xu2025qwen2} substantially outperforms other multimodal alternatives.

This observation motivates our final model design. Specifically, we employ the Qwen3 text encoder and the CLIP~\cite{radford2021learning} image encoder to independently extract textual and visual representations, which are then jointly fused and compressed through a dedicated fusion module composed of Q-Former layers and MLP layers. During the compression process, the query tokens in the Q-Former are progressively refined and enhanced via residual initialization across layers. Compared with simply replacing multimodal backbones, focusing on the fusion architecture itself yields more than a $30\%$ relative improvement in HitRate@1 over the initial OneRec-style variants, while achieving up to a 32× speedup in embedding inference.

\paragraph{Semantic ID Evolution.}
We investigate the impact of different item representation strategies by iterating through four versions of semantic IDs. The progression moves from a unimodal textual baseline to a sophisticated multi-view contrastive alignment framework. Table~\ref{tab:sid_versions_transposed} presents the quantitative evaluation of these SID versions. We assess the quality of the learned representations using Codebook Coverage, Semantic Diversity, P99 Collision (SKU count at P99).
\begin{table}[h]
\centering
\caption{Detailed Evaluation of Semantic ID Versions.}
\label{tab:sid_versions_transposed}
\begin{tabular}{llcccc}
\toprule
\multicolumn{2}{c}{\textbf{Metric}} & \textbf{V1 (Textual)} & \textbf{V2 (MiniCPM)} & \textbf{V3 (Fusion)} & \textbf{V4 (Multi-Source)} \\ 
\midrule
\multirow{3}{*}{\textbf{Codebook Coverage} ($\uparrow$)} & Codebook1 & 100\% & 100\% & 100\% & \textbf{100\%} \\
 & Codebook2 & 72.33\% & 31.22\% & \textbf{70.03\%} & 69.56\% \\
 & Codebook3 & 2.74\% & 0.013\% & 0.14\% & \textbf{0.15\%} \\
\midrule
\multirow{3}{*}{\textbf{Cluster Purity} ($\uparrow$)} & Cate1 & 86.30\% & 83.31\% & 84.38\% & \textbf{92.80\%} \\
 & Cate2 & 73.45\% & 66.80\% & 70.35\% & \textbf{79.90\%} \\
 & Cate3 & 53.68\% & 41.99\% & 48.41\% & \textbf{59.73\%} \\
\midrule
\multirow{3}{*}{\textbf{SID Collision} ($\downarrow$)} & P90 & 10 & 6 & 2 & \textbf{2} \\
 & P99 & 79 & 21 & 9 & \textbf{9} \\
 & P999 & 419 & 73 & \textbf{34} & 35 \\
\midrule
\multirow{3}{*}{\textbf{Load Balance} ($\to 100\%$)} & P25 & 32.49\% & 88.53\% & 89.25\% & \textbf{89.92\%} \\
 & P75 & 144.22\% & 114.66\% & 107.39\% & \textbf{105.40\%} \\
 & P90 & 212.84\% & 122.14\% & 118.60\% & \textbf{118.09\%} \\
\bottomrule
\multicolumn{6}{l}{\footnotesize \textbf{Note:} $\uparrow$: Higher is better; $\downarrow$: Lower is better; $\to 100\%$: Closer to 100\% is better.} \\
\end{tabular}%
\end{table}

The detailed configurations for each version are as follows:

\begin{itemize}
    \item \textbf{V1 Semantic ID (Textual Baseline):} 
    This version constructs discrete latent representations purely from textual semantics. Item embeddings are encoded using the pre-trained M3E~\cite{M3E} based on item titles. These continuous embeddings are then discretized via RQ-KMeans into a hierarchical 3-layer codebook, with a vocabulary size of 2,048 per level.
    
    \item \textbf{V2 Semantic ID (Multimodal \& Behavioral):} 
    This iteration introduces multimodal understanding and behavioral alignment. We utilize the MiniCPM-V-8B to encode both item titles and images. To bridge the semantic gap, we apply contrastive learning where positive pairs are derived from I2I co-occurrence graphs mined via the Swing algorithm. The resulting embeddings are quantized into a larger hierarchical codebook (3 layers, 8,192 width).
    
    \item \textbf{V3 Semantic ID (Advanced Fusion Architecture):} 
    This version upgrades the backbone to a powerful fusion architecture combining Qwen3 and CLIP~\cite{radford2021learning}. The model is trained using the previously described fusion structure with a contrastive objective, strictly supervising the learning process using swing-based I2I pairs as ground-truth labels. The latent space is mapped to a 3-layer codebook with a width of 8,192 using RQ-KMeans.

    \item \textbf{V4 Semantic ID (Multi-Source Alignment):}
    The most advanced version leverages the Qwen3 and CLIP-ViT~\cite{radford2021learning} fusion backbone but significantly enhances the contrastive learning paradigm through multi-source supervision. We integrate diverse alignment signals, including Q2I associations, Swing-based behavioral I2I pairs, and semantic peers (items under the same query and category). This multi-view approach ensures robust semantic consistency before the final RQ-KMeans quantization (3 layers, 8,192 width).

\end{itemize}

\textbf{Analysis of Semantic ID Quality.} As shown in Table~\ref{tab:sid_versions_transposed}, the evolution of semantic ID versions demonstrates a clear trajectory towards higher expressiveness and stability. While the V1 (Textual) baseline appears to have high codebook coverage, this is primarily an artifact of its significantly smaller codebook size ($2048$ vs. $8192$), which fundamentally limits its resolution capacity---evidenced by its poor P999 collision count of 419. In contrast, V2 (MiniCPM) suffers from low coverage in deeper levels (0.013\% at L3), indicating partial codebook collapse. The V3 (Fusion) and V4 (Multi-Source) iterations successfully address these issues by leveraging robust multimodal alignment. Most notably, V4 achieves state-of-the-art performance across all critical dimensions: it attains the highest \textit{Cluster Purity} (92.80\% at Cate-L1) by injecting explicit category semantics; delivers superior \textit{SID Collision} (reducing P999 collisions to just 35, ensuring fine-grained item distinction); and maintains an optimal \textit{Load Balance} (P90 of 118.09\%), confirming that the multi-source contrastive learning paradigm effectively maximizes the utilization and uniformity of the high-dimensional latent space.

\paragraph{Generative Backbone Architecture Ablation.}
Having established the optimal semantic ID representation (V4), we further investigate the scaling laws and architectural hyperparameters of the generative backbone. We construct several pre-trained models with varying capacities by systematically adjusting the depth of encoder/decoder layers, the model dimensions (hidden state and intermediate sizes), and the Mixture-of-Experts (MoE) configurations (total vs. active experts)~\cite{sha1345}. All models are trained on the same dataset to isolate the effects of parameter scaling.

Table~\ref{tab:model_configs} details the architectural specifications of the evaluated model variants, scaling from 0.1B to 3.0B parameters.

\begin{table}[h]
\centering
\caption{Model Configurations for Generative Backbone Ablation.}
\label{tab:model_configs}
\begin{tabular}{ccccc}
\toprule
\textbf{Model Size (Tot/Act)} & \textbf{Enc Layers} & \textbf{Dec Layers} & \textbf{Dim (Hidden/Inter)}  & \textbf{Experts (Tot/Act)} \\ 
\midrule
0.1B / 0.1B & 4 & 4 & 1024/512 & 2/1 \\
0.4B / 0.3B & 4 & 6 & 2048/1024 & 4/2 \\
0.7B / 0.4B & 4 & 8 & 2048/1024 & 8/2 \\
1.5B / 0.4B & 4 & 8 & 2048/1024 & 24/2 \\
3.0B / 0.6B & 4 & 16 & 2048/1024 & 24/2 \\
\bottomrule
\end{tabular}%
\end{table}

\begin{figure}[h]
    \centering

    \includegraphics[width=0.8\linewidth]{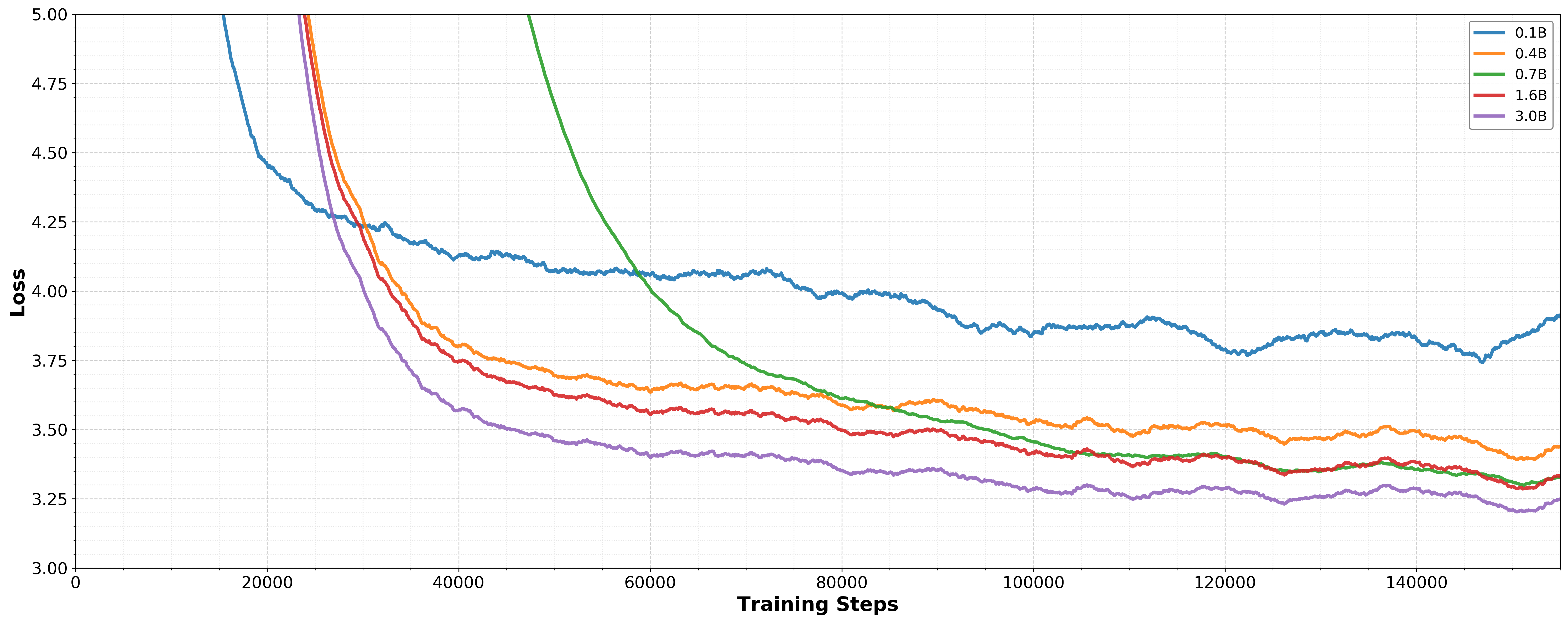}
    \caption{NTP Loss Scaling Laws.}
    \label{fig:scaling_law}
\end{figure}

We assess the generative performance of these variants using hitrate and recall metrics. Furthermore, to analyze training dynamics, we monitor the NTP Loss. Table~\ref{tab:model_performance} and Figure~\ref{fig:scaling_law} present the retrieval metrics and the loss scaling laws, respectively.

\begin{table}[h]
    \centering
    \caption{Performance Comparison.}
    \label{tab:model_performance}
        \begin{tabular}{ccccc}
        \toprule
        \textbf{Model Size (Tot)} & \textbf{HR@1} & \textbf{HR@10} & \textbf{Recall@10} & \textbf{Recall@30} \\ 
        \midrule
        0.1B & 3.99\% & 13.17\% & 10.10\% & 15.11\% \\
        0.4B & 4.42\% & 15.03\% & 11.38\% & 17.34\% \\
        0.7B & 4.84\% & 16.33\% & 12.32\% & 18.71\% \\
        1.6B & 4.92\% & 16.61\% & 12.51\% & 19.01\% \\
        3.0B & \textbf{5.02\%} & \textbf{16.99\%} & \textbf{12.78\%} & \textbf{19.53\%} \\
        \bottomrule
        \multicolumn{5}{l}{\footnotesize \textbf{Note:} Results represent overall metrics across nearly all core scenarios at JD App using the full-scale training dataset.}
        \end{tabular}%
\end{table}

\textbf{Scaling Law Validation and MoE Saturation Analysis.} 
The experimental results in Table~\ref{tab:model_performance} and Figure~\ref{fig:scaling_law} empirically validate the scaling laws within our generative recommendation framework. We observe a general positive correlation between model capacity and retrieval performance. Specifically, scaling the model from 0.1B to 3.0B yields a monotonic improvement in HR@10, rising from 13.17\% to 16.99\%. Figure~\ref{fig:scaling_law} demonstrates this, illustrating that larger models generally achieve lower asymptotic NTP loss.

A detailed analysis reveals a distinct plateauing phase between the 0.7B and 1.6B variants, where the loss curves exhibit significant overlap despite the increase in total parameter count. We speculate that this diminishing return might be linked to the sparse activation mechanism of the MoE architecture. As both configurations utilize the same number of \textit{active experts} (2) per token, the effective parameter count involved in each forward pass remains comparable, potentially limiting the immediate gains from simply expanding the expert pool. However, the \textit{3.0B model} appears to overcome this bottleneck, likely due to the increased depth (16 decoder layers), which extends the computational path and further drives down the NTP loss.

\subsection{Instruction-Following Unification Analysis}

This section investigates the efficacy of our proposed instruction-following mechanism, which serves as the cornerstone for unifying disparate recommendation tasks into a single generative framework. In our architecture, a composite \textit{Instruction Token} is prepended to the decoder input to steer the generation process. It is worth noting that to enable rapid and efficient exploration of model structures, the ablation studies in this subsection are conducted on a representative subset of the full dataset. To rigorously validate this design, we conduct a five-stage analysis: first, identifying the optimal \textit{positional strategy} for integrating the token; second, dissecting the contribution of its \textit{constituent components} (Scenario vs. Trigger); third, evaluating the impact of the IGR components; fourth, demonstrating the \textit{superiority of the unified model} over traditional isolated SFT baselines across five core scenarios; and finally, verifying the \textit{sensitivity} of the decoder to these instructions via masking experiments.

\paragraph{Instruction Token Integration Strategies.}
We first determine the optimal position for inserting the instruction token into the decoder's input sequence relative to the Begin-Of-Sequence (BOS) marker. We compare five strategies: (1) \textbf{No Instruction} (Baseline); (2) \textbf{Replace BOS}; (3) \textbf{Add to BOS} (element-wise sum); (4) \textbf{Insert Left of BOS}; and (5) \textbf{Insert Right of BOS}.

Table~\ref{tab:instruction_strategies} presents the comparative results. The ``Insert Right of BOS'' strategy consistently yields the highest hitrate and recall across all metrics. This finding suggests that placing the instruction \textit{after} the standard BOS marker allows the decoder to first initialize its state and then immediately condition its subsequent autoregressive generation on the specific context, providing the most effective guidance flow.

\begin{table}[h]
\centering
\small  
\caption{Performance comparison of different instruction token integration strategies.}
\label{tab:instruction_strategies}
\begin{tabular}{lcccc}
\toprule
\textbf{Integration Strategy} & \textbf{HR@1} & \textbf{HR@10} & \textbf{Recall@10} & \textbf{Recall@30} \\ 
\midrule
No Instruction & 2.78\% & 10.38\% & 8.18\% & 13.01\% \\
Replace BOS & 3.30\% & 12.08\% & 9.12\% & 14.38\% \\
Add to BOS & 3.50\% & 12.59\% & 9.52\% & \textbf{14.93\%} \\
Insert Left of BOS & 3.33\% & 12.17\% & 9.21\% & 14.50\% \\
\textbf{Insert Right of BOS} & \textbf{3.53\%} & \textbf{12.68\%} & \textbf{9.58\%} & 14.91\% \\
\bottomrule
\end{tabular}
\end{table}

\paragraph{Instruction Component Ablation.}
Building upon the optimal ``Insert Right of BOS'' strategy, we further dissect the composition of the instruction token itself. We verify whether fusing both \textit{Scenario ID} and \textit{Trigger Item ID} is necessary compared to using either component in isolation.

As shown in Table~\ref{tab:instruction_composition}, while using Scenario ID or Trigger Item ID alone improves performance over the ``No Instruction'' baseline, the \textit{Concatenated Instruction} (Scenario + Trigger) achieves better results. Furthermore, the \textit{Fused Instruction} approach, which projects the combined features into a unified representation before integration, delivers the best performance across all metrics. This indicates that the two signals provide complementary guidance: the Scenario ID defines the global domain characteristics (e.g., price sensitivity in homepage vs. add-on logic in cart), while the Trigger Item ID provides fine-grained, localized user intent context. The superior performance of the fused approach suggests that a deeper integration of these signals allows the model to better capture the interactions between scenario context and user intent.

\begin{table}[h]
\centering
\small
\caption{Ablation study on Instruction Token components.}
\label{tab:instruction_composition}
\begin{tabular}{lcccc}
\toprule
\textbf{Instruction Components} & \textbf{HR@1} & \textbf{HR@10} & \textbf{Recall@10} & \textbf{Recall@30} \\ 
\midrule
No Instruction (Baseline) & 2.78\% & 10.38\% & 8.18\% & 13.01\% \\
Scenario ID Only (1 token) & 3.30\% & 12.17\% & 9.22\% & 14.50\% \\
Trigger Item ID Only (1 token) & 3.22\% & 11.60\% & 9.13\% & 13.98\% \\
Concatenated (Scenario + Trigger, 2 tokens) & 3.53\% & 12.68\% & 9.58\% & 14.91\% \\
\textbf{Fused (Scenario + Trigger, 1 token)} & \textbf{3.60\%} & \textbf{12.82\%} & \textbf{9.68\%} & \textbf{15.08\%} \\
\bottomrule
\end{tabular}
\end{table}

\paragraph{IGR Ablation.}
To validate the effectiveness of IGR mechanism and Q2I alignment, we conduct ablation studies under search-dominated scenarios. As shown in Table~\ref{tab:q2i_igr_ablation}, introducing IGR alone improves the retrieval quality by better focusing on relevant historical interactions. The full model with both IGR and Q2I alignment achieves the best performance, demonstrating that explicit alignment between query and item spaces is crucial for effective IGR.

\begin{table}[h]
\centering
\small 
\caption{Ablation study on IGR components}
\label{tab:q2i_igr_ablation}
\begin{tabular}{lcccc}
\toprule
\textbf{Configuration} & \textbf{HR@1} & \textbf{HR@10} & \textbf{Recall@10} & \textbf{Recall@30} \\ 
\midrule
Base Model (w/o IGR/Q2I) & 3.76\% & 12.20\% & 9.87\% & 15.53\% \\
+ IGR Only & 4.02\% & 12.91\% & 10.25\% & 15.95\% \\
+ IGR+Q2I (Full) & \textbf{4.19\%} & \textbf{13.38\%} & \textbf{10.52\%} & \textbf{16.23\%} \\
\bottomrule
\end{tabular}
\end{table}

\paragraph{Unified Instruction-Following Model vs. Independent SFT Baselines.}
We compare our \textbf{Unified Instruction-Following Model} against the industry-standard \textbf{Pretrain and Scenario Independent SFT} approach, where separate models are fine-tuned exclusively for each specific scenario.
Table~\ref{tab:unified_vs_sft} illustrates the performance across five distinct deployment scenarios. The unified model consistently outperforms the isolated baselines. This superiority is attributed to \textit{Synergistic Knowledge Transfer}, where high-resource Scenarios (like Homepage) enhance the representation learning for lower-resource Scenarios; \textit{Universal User Modeling}, which captures a holistic view of user interests; and significantly improved \textit{Operational Efficiency}, as a single model reduces maintenance and GPU overhead compared to managing five separate checkpoints.

\begin{table}[h]
\centering
\footnotesize 
\setlength{\tabcolsep}{3.5pt} 
\caption{Performance comparison: Unified Instruction-Following Model vs. Independent SFT Baselines across six core scenarios.}
\label{tab:unified_vs_sft}
\begin{tabular}{llcccccc}
\toprule
\textbf{Metric} & \textbf{Model Type} & \textbf{Scenario 1} & \textbf{Scenario 2} & \textbf{Scenario 3} & \textbf{Scenario 4} & \textbf{Scenario 5} & \textbf{Scenario 6} \\ 
\midrule
\multirow{2}{*}{\textbf{HR@1}} 
 & Independent SFT & 6.39\% & 8.17\% & 1.12\% & 1.83\% & 7.22\% & 5.29\% \\
 & \textbf{Unified Model} & \textbf{15.39\%} & \textbf{20.75\%} & \textbf{17.24\%} & \textbf{6.34\%} & \textbf{10.54\%} & \textbf{25.75\%} \\ 
\midrule
\multirow{2}{*}{\textbf{HR@10}} 
 & Independent SFT & 23.29\% & 29.05\% & 5.22\% & 8.44\% & 29.84\% & 19.38\% \\
 & \textbf{Unified Model} & \textbf{46.73\%} & \textbf{55.02\%} & \textbf{53.57\%} & \textbf{29.89\%} & \textbf{37.90\%} & \textbf{62.62\%} \\ 
\bottomrule
\end{tabular}
\end{table}

\paragraph{Trigger Instruction Sensitivity Analysis.}
To rigorously verify that the unified model effectively utilizes the \textit{Trigger Item} component of the instruction token as a control signal---rather than merely ignoring it---we conduct an inference-time sensitivity analysis. 
We focus specifically on Scenarios 3, 4, 5, and 6, which are explicitly designed to be driven by trigger items (Scenarios 1 and 2 are excluded as they operate without specific trigger inputs).

In this experiment, we freeze the model parameters and replace the authentic \textit{Trigger Item ID} in the instruction token with a generic ``Default'' embedding during the decoding process. As presented in Table~\ref{tab:trigger_sensitivity}, substituting the specific trigger with a default value leads to a substantial degradation in retrieval metrics across all tested scenarios. This sharp drop confirms the validity of our instruction mechanism: the decoder heavily relies on the fine-grained trigger signal to effectively contextualize the user's immediate intent, acting as a critical ``steering wheel'' for the generative process.

\begin{table}[h]
\centering
\footnotesize
\setlength{\tabcolsep}{6pt} 
\caption{Sensitivity Analysis: Impact of masking the Trigger Item ID during inference (Scenarios 3, 4, 5, 6).}
\label{tab:trigger_sensitivity}
\begin{tabular}{llcccc}
\toprule
\textbf{Metric} & \textbf{Inference Setting} & \textbf{Scenario 3} & \textbf{Scenario 4} & \textbf{Scenario 5} & \textbf{Scenario 6} \\ 
\midrule
\multirow{2}{*}{\textbf{HR@1}} 
 & \textbf{Correct Trigger (Instruction)} & \textbf{20.75\%} & \textbf{6.34\%} & \textbf{10.54\%} & \textbf{25.75\%} \\
 & Masked Trigger (Default) & 10.71\% & 1.96\% & 9.26\% & 18.55\% \\ 
\midrule
\multirow{2}{*}{\textbf{HR@10}} 
 & \textbf{Correct Trigger (Instruction)} & \textbf{55.02\%} & \textbf{29.89\%} & \textbf{37.90\%} & \textbf{62.62\%} \\
 & Masked Trigger (Default) & 40.96\% & 11.31\% & 35.74\% & 50.52\% \\ 
\bottomrule
\end{tabular}
\end{table}

\subsection{Post-training of OxygenREC}

In this section, we primarily validate the effectiveness of synthetic data and the results of training a unified model across various scenarios. OneRec-v2~\cite{zhou2025onerecv2} proposes an online RL learning approach based on user behavior data streams. Through interaction with extensive real-time user data streams, it updates policies according to real user behavior preferences. However, this approach heavily relies on real-time user feedback and often suffers from reward hacking in data-sparse scenarios. We address this by adopting a post-training method that generates a large amount of synthetic data and performs unified post-training process to enhance the model's robustness across different scenarios.

\begin{figure}[h]
    \centering
    \includegraphics[width=0.65\textwidth]{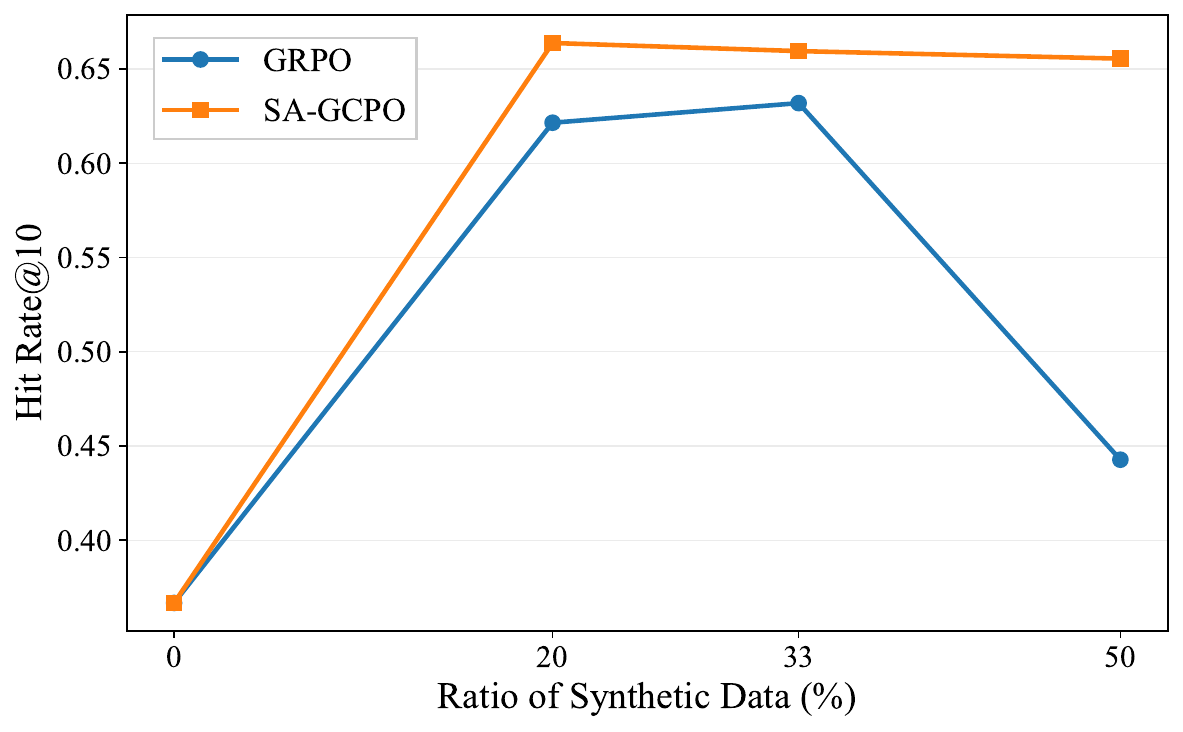} 
    \caption{Evaluation of synthetic data}
    \label{fig:synthetic}
\end{figure}

\paragraph{Validate the effectiveness of synthetic data.}
Firstly, we conducted experiments to validate the effectiveness of synthetic data. Specifically, we utilized user profiles from historical data and candidate items generated by the policy model to query the ranking model to obtain reward scores. These scores were then employed as rewards for policy learning.

We employ the pre-trained OxygenREC-0.7B MOE as the backbone model. Then we conduct the RL training to validate the effectiveness of synthetic data. As illustrated in Figure~\ref{fig:synthetic}, we select one of the scenarios to present HR@10 across varying proportions of synthetic data used during training. The proportion denotes the percentage of synthetic samples incorporated into the training dataset. When the ratio of synthetic data is 0, it represents the performance of the warm-start pretrained model at the test data. We observe that the performance of GRPO exhibits considerable instability across varying proportions of synthetic data, whereas SA-GCPO demonstrates more consistent results. Moreover, SA-GCPO shows superior performance compared to GRPO across different proportions of synthetic data, which demonstrates the effectiveness of our proposed method. 

\begin{table}[h]
  \caption{Evaluation of proposed SA-GCPO with other methods }
  \label{tab:grpo}
  \centering
  \begin{tabular}{llll}
    \toprule
    \textbf{Methods} & \textbf{Ratio of synthetic data}   & \textbf{HR@1} & \textbf{HR@10}\\
    \midrule
    OxygenREC0.7B-GRPO   & 33\% & 23.85\%  &  62.15\%\\
    OxygenREC0.7B-GSPO   & 33\% & 24.13\%  &  62.88\%\\
    \textbf{OxygenREC0.7B-SA-GCPO}   & \textbf{33\%} &  \textbf{25.58\%} & \textbf{65.95\%} \\
    \bottomrule
  \end{tabular}
\end{table}

\paragraph{Effectiveness of proposed SA-GCPO.}
When it comes to the post-training stage, there are numerous effective methods to enhance the effectiveness of RL stage. GRPO~\cite{guo2025deepseek} has been widely applied in post-training across LLMs, VLMs and GRs. Recently, many modified approaches based on GRPO have emerged, such as GSPO~\cite{zheng2025group}, which enhances model performance by computing importance sampling along the sequence, which also represents a promising attempt in the field of GR. This is because treating semantic IDs as an entire sequence for importance sampling is more meaningful than performing importance sampling at the token level. We implement this approach by training in our OxygenREC framework. To validate the effectiveness of our proposed SA-GCPO, we conducted experiments comparing its performance with that of GRPO and modified version of GSPO. We warm-start pre-trained OxygenREC-0.7B as the backbone and control the proportion of synthetic data at $33\%$. We show the experimental result in Table~\ref{tab:grpo}. As it shows in the table, GSPO is slightly better than that of GRPO in terms of HR@1 and HR@10. SA-GCPO achieves +1.45pp and +1.73pp compared with the HR@1 of GSPO and GRPO, respectively. When it comes to HR@10, SA-GCPO achieves more than +3pp, which demonstrates the superior performance of our RL method. 

\begin{table}
  \caption{Ablation study of temperatures set for positive and negative samples of SA-GCPO  }
  \label{tab:temp}
  \centering
  \begin{tabular}{llll}
    \toprule
     \textbf{$\tau_{\text{pos}}$}     & \textbf{$\tau_{\text{neg}}$}   & \textbf{HR@1} & \textbf{HR@10}\\
    \midrule
     1.0   & 1.05 & 25.35\%  &  65.64\%\\
     1.0   & 1.0 & 25.48\% &  65.95\%\\
    \textbf{1.0}   & \textbf{0.95} &  \textbf{25.51\%} & \textbf{66.01\%} \\
    \bottomrule
  \end{tabular}
\end{table}

\paragraph{Ablation study for different settings of $\tau_{\text{pos}}$ and $\tau_{\text{neg}}$. } In addition, we conducted ablation experiments to validate the effect of separate temperature setting for positive and negative advantage samples in our proposed SA-GCPO. As shown in Table~\ref{tab:temp}, we experimented with different temperature coefficients for positive and negative samples. When $\tau_{\text{pos}} > \tau_{\text{neg}}$, the model performance becomes more stable. Assigning a lower temperature coefficient to negative samples helps prevent performance collapse during RL training and enhances overall training stability.

\subsection{Online A/B Test Performance and Industrial Impact}

\textbf{Deployment Scenarios and User Lifecycle Coverage.}
The implementation of the xLLM inference engine significantly enhanced the model's throughput and latency performance, enabling the generative recommendation system to handle high-concurrency requests with reduced computational overhead. This performance breakthrough facilitated the deployment of our model across three sequentially dependent scenarios on the JD App, forming a holistic closed-loop covering the user's entire session lifecycle.
\begin{itemize}
    \item \textbf{Phase 1: Interest Triggering (Homepage Floor).} The first deployment covers the high-traffic Scenario 1\footnote{Scenario 1: A ``Billion Subsidy'' program where the platform subsidizes brand-name products to offer users price advantages below market cost.} and Scenario 2\footnote{Scenario 2: A ``Budget \& Free Shipping'' channel designed to satisfy user demand for the lowest price range within specific categories and specifications.} on the JD App homepage. Characterized by massive DAU and strict low-latency constraints, this scenario serves as the entry point. It recommends visually engaging items highly relevant to user interests, attracting clicks that guide users into the downstream feeds.
    
    \item \textbf{Phase 2: Deep Exploration (Feeds Recommendation).} When users click on SKUs in the two scenarios (Scenario 1 and Scenario 2) on the homepage, they will be directed to the corresponding feed pages (Scenario 3 and Scenario 4). Here, the system generates recommendations based on the ``trigger SKU'' (the clicked item) and the user's long/short-term behavior. This scenario encourages prolonged engagement (browsing duration) through an infinite scroll mechanism, fostering intent for adding-to-cart or purchasing.
    
    \item \textbf{Phase 3: Immediate Conversion (Checkout Path Recommendations).} The final phase targets the transaction process. When a user adds an item to the cart, an \textit{Add-to-Cart Overlay (Scenario 5)} immediately exposes a sequence of related items. Similarly, the \textit{Checkout Add-on (Scenario 6)} page displays recommendations during the payment process, allowing users to conveniently pick up supplementary items. These Checkout Path scenarios capitalize on the user's strong purchase intent to drive immediate additional conversions.
\end{itemize}
Together, these scenarios create a synergistic pathway: attracting attention (Guidance), retaining interest (Retention), and facilitating transactions (Conversion), thereby optimizing the user experience throughout their entire request lifecycle.

\paragraph{Online A/B Testing Results.}
We conducted rigorous online A/B testing across the aforementioned scenarios to validate the business impact of our generative framework. For each deployment, users were randomly bucketed into experimental and control groups, each accounting for 10\% of the total traffic. The experimental group was served by our proposed model, while the control group relied on the existing production baseline. 

As summarized in Table~\ref{tab:online_ab_results}, the generative model achieved statistically significant improvements across all key business metrics---including User Click-Through Rate (UCTR), User Conversion Rate (UCTCVR), Order Volume, and Gross Merchandise Value (GMV)---demonstrating its robustness across diverse operational environments ranging from high-traffic homepage slots to high-intent transactional flows. The table also details the specific operational characteristics (DAU, Latency, Candidate Scale) for each scenario.

\begin{table}[h]
\centering
\caption{Online A/B Test Lift at First Launch}
\label{tab:online_ab_results}
\begin{tabular}{llccccc}
\toprule
\multicolumn{2}{c}{\textbf{Scenario}} & \textbf{UCTR} & \textbf{UCTCVR} & \textbf{Order Volume} & \textbf{GMV} & \textbf{Latency} \\ 
\midrule
\multirow{2}{*}{\textbf{Homepage Floor}} & Scenario 1 & +0.68\% & +2.71\% & +2.81\% & +4.52\% & \multirow{2}{*}{50ms} \\
 & Scenario 2 & +3.55\% & +2.26\% & +2.21\% & +8.40\% &  \\ 
\midrule
\multirow{2}{*}{\textbf{Channel Feeds}} & Scenario 3$^{*}$ & -0.25\% & +7.89\% & +8.03\% & +1.46\% & \multirow{2}{*}{80ms} \\
 & Scenario 4 & +0.78\% & +2.17\% & +1.49\% & +1.66\% &  \\ 
\midrule
\multirow{2}{*}{\textbf{Checkout Path}} & Scenario 5 & +0.40\% & +4.21\% & +4.28\% & +11.80\%  & \multirow{2}{*}{50ms} \\
 & Scenario 6 & +3.29\% & +3.00\% & +2.92\% & +4.15\% &  \\ 
\bottomrule
\end{tabular}
\par
\vspace{0.5ex}
\footnotesize
\raggedright
\textit{Note:} $^{*}$ denotes scenarios where \textbf{OxygenREC} has achieved end-to-end replacement of traditional recommendation systems, with the generative model replacing Recall and Pre-Ranking stages, while the Ranking stage serves as the  reward mapping service.
\end{table}

In summary, the online A/B test results affirm the superiority of the generative paradigm in real-world industrial settings. As evidenced in Table~\ref{tab:online_ab_results}, the proposed model delivers consistent performance uplifts across disparate traffic environments. Notably, in the Feeds scenarios (e.g., Scenario 3), the model drives a remarkable \textbf{8\%+} surge in order volume, indicating its ability to prioritize high-quality items that genuinely align with purchase intent rather than merely optimizing for shallow clicks. Furthermore, in the transaction-critical Checkout Path (Scenario 5-6), the model successfully capitalizes on immediate user intent to achieve substantial gains in order volume despite billion-scale candidate spaces. These results validate that the unified generative framework effectively translates semantic understanding into tangible business growth across the user's entire shopping lifecycle.

\section{Conclusion and Future Work}
In this work, we introduce \textbf{OxygenREC}, a generative recommendation system that successfully integrates deep reasoning capabilities with the strict latency and scalability requirements of real-world deployment. We implement and validated the effectiveness of a \textit{Fast--Slow Thinking} architecture: the ``slow'' part uses a near-line LLM to generate clear and high-quality \textit{Contextual Reasoning Instructions} by analyzing user behavior and context, while the ``fast'' part---a lightweight encoder-decoder model---generates recommendations in real-time based on those instructions. More importantly, we address the gap between training and deployment by ensuring consistency between the training data for the slow path (derived from historical search logs) and the data encountered online by the fast path.
It achieves smooth knowledge transfer without calling LLMs online---thereby maintaining low latency while still enabling intelligent reasoning.

Furthermore, we explored \textbf{instruction-following in GR}, a core capability of LLM, which has been long overlooked in recommendation tasks. We achieved effective instruction control through semantic alignment. Our \textit{Instruction-Guided Retrieval} module retrieves the most relevant historical user behaviors based on the instructions, and our \textit{Query-to-Item} loss function ensures consistency between the recommendation results and the intent of the instructions, so that the final output truly reflects the intended reasoning logic. It provides a valuable reference paradigm for industrial‑scale recommendation models.

Finally, OxygenREC embodies the concept of \textbf{`train-once-deploy-everywhere'}. Unlike training separate models for different scenarios (such as the homepage or cart), we convert scenario-specific information into structured \textit{Scenario Instructions}. We enable a single model to learn to handle tasks across all scenarios through shared reward signals using \textit{Soft Adaptive Group Clip Policy Optimization}. The system has been fully deployed in JD.com’s core recommendation scenarios. OxygenREC delivered significant business gains from its initial launch across multiple scenarios, demonstrating its real-world impact and practical value. Future work includes the following promising directions.

\paragraph{Latency Optimization through Non-Autoregressive Generation.}
The current framework, reliant on sequential NTP, faces a fundamental scalability barrier: decoding latency increases linearly with the required recommendation list length, severely hindering high-throughput real-time deployment. Our long-term goal is to overcome this limitation by transitioning to a highly efficient Non-Autoregressive (NAR) parallel generation paradigm.

This initiative aims to drastically minimize serving latency and maximize throughput by generating the entire sequence of semantic identifiers simultaneously. This shift is critical for maintaining performance as the model complexity and knowledge integration depth increase.

\paragraph{Multi-Scenario User Trajectory Modeling for Deep Intent Discovery.}
The current instruction system effectively utilizes immediate context and scene information. However, users' true purchase intent is often a complex decision trajectory spanning multiple distinct scenarios (e.g., Homepage, Search, Cart, Checkout) within the platform. Our focus will pivot towards \textit{multi-scenario user trajectory modeling} to capture the full context of user behavior.

This direction will involve integrating and analyzing cross-scenario sequences to uncover deep-seated user goals and the evolution of their intent between different pages. By capturing these complex dynamics, we can upgrade our instruction system to utilize richer, hierarchical intent signals for the LLM backbone, ensuring more precise and long-term optimal recommendations. This strategy will be further aligned with long-term user value via a robust closed-loop learning mechanism.


\clearpage

\bibliographystyle{plain} 
\bibliography{references} 
\clearpage

\appendix

\section{Related Work}
\label{app:related_work}

We briefly review related work along three axes: generative recommendation, LLM-based recommendation systems, and multi-scenario adaptation.

\paragraph{Generative Recommendation.}
GR has recently emerged as a paradigm shift, reformulating recommendation as an end-to-end sequence generation task~\cite{lin2025survey,zhou2025onerec,rajput2023recommender}. Unlike traditional discriminative models that score candidates from a fixed pool, GR models directly generate target item identifiers. A core challenge in this paradigm lies in \textit{tokenization}---representing items as discrete tokens. Early works utilized raw textual IDs, while recent advancements have shifted towards SIDs~\cite{hui2025semantics,tan2024idgenrec,subbiah2025efficient,huang2025q}. Methods like RQ-VAE~\cite{rajput2023recommender} and residual quantization~\cite{gu2024residual} create hierarchical codes that capture semantic relationships, balancing expressiveness with vocabulary size. Recent works have explored unifying search and recommendation through generative approaches, like SynerGen~\cite{hou2025synergen} and IntSR~\cite{yan2025intsr}. However, these methods are built upon traditional ranking models rather than semantic ID-based generation. While some approaches integrate large language models for enhanced reasoning capabilities, like RecGPT series~\cite{bao2025recgpt, bao2025recgptv2}, they employ offline LLM inference for reasoning and rely on simple ANN-based dual-tower retrieval for online serving, which limits their ability to fully leverage LLM reasoning capabilities in real-time scenarios. From an architectural perspective, the backbone of GR typically employs transformer-based encoder--decoder or decoder-only architectures~\cite{vaswani2017attention,mei2023lightlm,xing2025reg4rec}. These unified structures are naturally aligned with modern GPU accelerators, offering significant throughput advantages over fragmented deep learning pipelines. However, despite these advances, most GR methods remain fundamentally \textit{inductive}. They excel at fitting historical interaction patterns but lack the \textit{deductive reasoning} capabilities required to infer complex user intents from sparse or novel contexts.

\paragraph{Large Language Models for Recommendation.}
The success of LLMs in NLP~\cite{brown2020language,achiam2023gpt,ouyang2022training,jaech2024openai} has spurred interest in adapting them for recommendation. Some approaches directly utilize LLMs as the recommendation \textit{backbone}~\cite{he2025plum,chen2024hllm,liu2025onerec}, leveraging their vast world knowledge and reasoning abilities. For instance, OneRec-think~\cite{liu2025onerec} integrates Chain-of-Thought (CoT) reasoning to enhance interpretability. However, deploying such models in industrial settings faces severe \textit{latency and cost constraints}. The computational overhead of autoregressive decoding with billion-scale parameters is often prohibitive for real-time serving~\cite{zhou2025onerecv2}.

\paragraph{Multi-Scenario Recommendation and Controllability.}
Industrial e-commerce platforms serve diverse scenarios (e.g., Homepage, Feeds, Cart), each with distinct user behaviors and objectives. Traditional Multi-Scenario Recommendation (MSR) approaches have primarily been developed for \textit{discriminative} ranking models, relying on specialized architectural interventions like gating~\cite{sheng2021one}, star-topology units~\cite{chang2023pepnet}, or routing mechanisms to mitigate \textit{negative transfer}. While effective, these methods introduce structural complexity and create information bottlenecks through implicit parameter modulation, often requiring maintaining separate model instances or intricate routing logic that increases operational and maintenance costs.

However, adapting these traditional MSR techniques to GR presents unique challenges. Most existing multi-scenario approaches for GR~\cite{hou2025synergen} remain largely exploratory and do not fully address the fundamental trade-off between unified modeling and scenario-specific adaptation. Related challenges also manifest in cross-domain recommendation~\cite{zhu2021cross} and handling imbalanced data distributions~\cite{buda2018systematic} across different scenarios, but solutions remain predominantly discriminative in nature.

\clearpage

\section{Contextual Reasoning Instructions }
\label{app:prompt_design}

The efficacy of instruction-following generative recommendation hinges critically on the quality of the reasoning instructions. To bridge raw behavioral signals with high-level, interpretable intent, we design a structured prompting strategy that synthesizes spatiotemporal context, user profiles, and multi-scale behavioral dynamics into natural-language instructions. This enables the LLM to perform deductive reasoning grounded in real-world semantics rather than mere pattern matching.

\paragraph{Spatiotemporal and User Profile Instructions.}
We enrich prompts with three foundational contextual signals and generate instructions based on geography, time, and identity: \textit{spatial features} (geographic location such as city or climate zone, which reflects region-specific demand like winter gear in northern cities), \textit{temporal features} (calendar date, seasonality, holidays, and diurnal patterns such as lunchtime browsing vs. evening shopping, capturing time-sensitive intent), and \textit{user profile features} (demographic and device attributes---including age group, gender, consumption tier, device type, and registered region---that shape baseline preferences).

\begin{tcolorbox}[
    breakable,
    enhanced,
    colframe=black!50,
    colback=gray!5,
    boxrule=0.6pt,
    arc=4pt,
    title={Spatiotemporal Reasoning},
    fonttitle=\bfseries\small,
    top=6pt,
    bottom=6pt,
    left=8pt,
    right=8pt
]

\textbf{Input:} 
\begin{itemize}[leftmargin=*]
\item {Location}: Beijing 
\item {Date}: December 21, 2025 
\item {User Profile}: male, 30 years old, white-collar, mid-to-high spending tier 
\end{itemize}

\textbf{Output: (Reasoning  → Instructions)}

Based solely on geographic and temporal context:
\begin{itemize}[leftmargin=*,nosep]
    \item \textit{Temporal inference}: December 21 is the winter solstice, which is a traditional Chinese festival → ``Frozen dumplings (traditional winter solstice food)''
    \item \textit{Spatial inference}: Beijing’s dry, sub-zero winter climate implies strong need for thermal protection during outdoor activities. → ``Bosideng men's long down jacket 90\% white duck down''
\end{itemize}

\noindent

\vspace{0.3em}

\end{tcolorbox}

\paragraph{Prompt Design for User Intent Mining.}
We split the prompt design of the User Intent Mining module into two parts: input and system prompt. Input describes what information we provide to the model, while the system prompt specifies how the model should reason and what format it should output.

Input. The input contains two complementary behavior sequences, written as a time-ordered natural language behavior narrative. This helps the model capture sequential patterns (e.g., ``search → add to cart → browse category'') and infer the user’s short-term shopping goal: (1) Long-term behavior sequence: up to 90 days of past interactions (click, add-to-cart, purchase), which reflects stable category preferences and long-term interests; (2) Short-term behavior sequence: interactions within the last 24 hours, which reflect session-level intent and recent context changes.

System Prompt. The system prompt guides the model to extract key signals from user behaviors and controls the output format. We ask the model to consider factors such as multi-level category interest, intent strength of different action types, time decay to emphasize recent behaviors, and scenario-aware category grouping, and to write high-confidence signals explicitly into the generated Instruction and Reason.

\begin{tcolorbox}[
    breakable,
    enhanced,
    colframe=black!50,
    colback=gray!5,
    boxrule=0.6pt,
    arc=4pt,
    title={User Intent Reasoning},
    fonttitle=\bfseries\small,
    top=6pt,
    bottom=6pt,
    left=8pt,
    right=8pt
]

\textbf{Input:}
\begin{itemize}[leftmargin=*,nosep]
    \item Added a full-suspension mountain bike (soft-tail) to cart on Dec 19 at 18:30 (high intent, L2: Outdoor Cycling);
    \item Clicked Huawei Mate 70 and iPhone 16 Pro on Dec 20 during lunch break (comparative browsing, L1: Consumer Electronics);
    \item Long-term Interests Input: Digital products, men's fashion, cycling, outdoor sports.
\end{itemize}

\vspace{0.3em}
\noindent

\vspace{0.5em}
\noindent
\textbf{Output: (Reasoning → Instructions)}

\begin{itemize}[leftmargin=*,nosep]
    \item \textit{Complementarity}: Cart addition of a mountain bike → infer need for cycling accessories (gloves, balaclava). → ``Winter cycling gloves with touchscreen compatibility'',``Thermal balaclava for outdoor riding''
    \item \textit{Interest fusion}: Intersection of long-term interests in \textit{cycling} and \textit{digital products} → ``GoPro HERO13 Black for action recording''
    \item \textit{Flagship exploration}: Recent clicks on premium smartphones  suggest competitive alternatives across brands with key differentiators. → ``Latest flagship smartphones (OPPO Find X9 -- periscope zoom / Xiaomi 17 -- 120W HyperCharge / Huawei Pura X -- satellite communication / iPhone 17 -- iOS ecosystem)''
\end{itemize}

\vspace{0.3em}

\end{tcolorbox}

This case demonstrates seamless fusion of seasonal culture, geographic climate adaptation, long-term identity (tech/outdoor enthusiast), and short-term behavioral signals---enabling context-aware, multi-intent recommendation generation via our LLM-powered framework.

\section{Detailed Definitions of Scenario Instructions}
\label{app:scenario_instructions}

To provide a deeper understanding of how OxygenREC adapts to diverse industrial contexts, we detail the composition and operational logic of our scenario instructions. As described in Section~\ref{sec:dual_instruction}, this instruction acts as a strong inductive bias, consisting of a \textit{Scenario Identifier} and an optional \textit{Trigger Item}.

\paragraph{Scenario Identifier: Guiding the Generative Distribution.}
The scenario ID serves as a high-level control token that shifts the model's generative distribution to match the specific business objectives and item pools of a given recommendation surface. Table~\ref{tab:scenario_examples} illustrates typical scenarios in our system.

\begin{table}[h]
\centering
\small
\caption{Examples of Scenario Identifiers and their Business Objectives.}
\label{tab:scenario_examples}
\begin{tabular}{>{\raggedright\arraybackslash}p{0.22\textwidth}|>{\raggedright\arraybackslash}p{0.30\textwidth}|>{\raggedright\arraybackslash}p{0.40\textwidth}}
\toprule
\textbf{Scenario ID} & \textbf{User Intent Context} & \textbf{Business Objectives \& Distribution} \\
\midrule
\texttt{Homepage\_Feed} & Passive browsing, low specific intent. & \textbf{Exploration}: Diverse categories, high CTR items, broad interest coverage. \\
\midrule
\texttt{Channel\_Feed} & Active entry via a specific category/item. & \textbf{Thematic Consistency}: Items strictly within the channel's theme (e.g., Digital). \\
\midrule
\texttt{Cart\_CrossSell} & High purchase intent for a specific item. & \textbf{Conversion}: Complementary accessories or bundles (e.g., Phone $\to$ Case). \\
\bottomrule
\end{tabular}
\end{table}

\paragraph{Trigger Item: Providing Explicit Contextual Anchors.}
The trigger item is an explicit signal representing the user's immediate focus. Crucially, its influence is \textit{modulated by the Scenario ID}, leading to context-dependent recommendations even for the same trigger. We illustrate this interaction with a concrete example where the trigger item is an \textbf{iPhone 15}.

\paragraph{Case 1: Channel Feed Scenario (Discovery).}
\begin{itemize}
    \item \textbf{Context}: The user clicks on an iPhone 15 on the homepage to enter the ``Digital Channel''. 
    \item \textbf{Role of Trigger}: The iPhone 15 serves as a \textit{Category Anchor}.
    \item \textbf{Model Behavior}: The model generates a feed of \textit{similar but diverse} high-end digital products (e.g., iPads, High-end Android phones, Bluetooth speakers). The goal is to encourage continued browsing within the tech theme.
\end{itemize}

\paragraph{Case 2: Item to Item Relative Recommendation (Completion).}
\begin{itemize}
    \item \textbf{Context}: The user adds the iPhone 15 to their shopping cart.
    \item \textbf{Role of Trigger}: The iPhone 15 serves as a \textit{Complementary Anchor}.
    \item \textbf{Model Behavior}: The model strictly generates \textit{functional accessories} (e.g., USB-C cables, MagSafe cases, AirPods). Generating another phone here would be a negative user experience.
\end{itemize}

This comparison demonstrates why both signals are necessary: the Trigger Item provides the \textit{content}, while the Scenario ID dictates the \textit{relationship} (Similarity vs. Complementarity).

\end{document}